\documentclass[12pt]{article}
\usepackage{graphicx}
\usepackage{latexsym,amssymb}
\newcommand{\preprint}[1]{\begin{flushright}#1\end{flushright}}

\newcommand \bra[1]{\left< {#1} \,\right\vert}
\newcommand \ket[1]{\left\vert\, {#1} \, \right>}
\newcommand \braket[2]{\hbox{$\left< {#1} \,\vrule\, {#2} \right>$}}
\newcommand{\bea}{\begin{eqnarray}}
\newcommand{\eea}{\end{eqnarray}}
\newcommand{\simgt}{\hbox{ \raise3pt\hbox to 0pt{$>$}\raise-3pt\hbox{$\sim$} }}
\newcommand{\simlt}{\hbox{ \raise3pt\hbox to 0pt{$<$}\raise-3pt\hbox{$\sim$} }}

\newcommand{\clfn}{\setcounter{footnote}{0}}

\def\to{\rightarrow}

\begin{document}
\preprint{hep-ph/9910424\\TU-579\\October 1999}
\vspace*{3cm}
\begin{center}
  {\bf\large
Gauge dependence and matching procedure of a nonrelativistic\\
 QED/QCD boundstate formalism    }
  \\[5mm]
  {
    K.~Hasebe and Y. Sumino
    }
  \\[5mm]
  {\it
    Department of Physics, Tohoku University\\
    Sendai, 980-8578 Japan
    }
\end{center}
\vspace{3cm}
\begin{abstract}
A nonrelativistic boundstate formalism used
in contemporary calculations is investigated.
It is known that the effective Hamiltonian of the boundstate
system depends on the choice of gauge.
We obtain the transformation charge $Q$ of the Hamiltonian
for an arbitrary infinitesimal change of gauge, by which
gauge independence of the mass spectrum and gauge dependences of the
boundstate wave functions are dictated.
We give formal arguments based on the BRST symmetry 
supplemented by power countings of Coulomb singularities of diagrams.
For illustration:
(1)we calculate $Q$ up to ${\cal O}(1/c)$,
(2)we examine gauge dependences of diagrams for a decay of a $q\bar{q}$
boundstate up to ${\cal O}(1/c)$ and show that
cumbersome gauge cancellations can be circumvented by directly 
calculating $Q$.
As an application we point out that the present calculations of
top quark momentum distribution in the $t\bar{t}$ threshold
region are gauge dependent.
We also show possibilities for incorrect calculations of physical quantities
of boundstates when the on-shell matching procedure is employed.
We give a proof of a justification for the use of the
equation of motion to simplify the form of a local NRQCD Lagrangian.
The formalism developed in this 
work will provide useful cross checks in computations involving
NRQED/NRQCD boundstates.
\end{abstract}

\newpage
\section{Introduction}
\label{s1}

Recently there has been much progress in our theoretical
understanding of 
nonrelativistic QED and QCD (NRQED and NRQCD) boundstates such as
positronium, $\Upsilon$ and remnant of toponium boundstates.
Following the idea of nonrelativistic effective field theory 
proposed by Caswell and Lepage \cite{cl},
formalisms necessary for precise descriptions of these boundstates
have been developed significantly \cite{bbl}-\cite{renormalon}.
At the same time there appeared many 
new calculations of higher order corrections to 
physical quantities of both
the NRQED \cite{kn}-\cite{cmy} and NRQCD boundstates \cite{qcdpot}-\cite{kp}
(boundstate mass, decay width, production and decay cross sections,
etc.).
Despite these developments, a completely systematic 
formulation necessary for computations of
these physical quantities in perturbative expansions
has not been established yet.
Among current best technologies the
asymptotic expansion of Feynman diagrams \cite{bs} seems to be most suited 
for calculations of the physical quantities, in particular for 
fixed-order calculations.
In addition, effective field theories are powerful tools in order to
sum up various large logarithms originating from 
the widely separated scales inherent in the problems.
Efficiencies and correctness of various effective theories are, however, 
still under debate.
(See Refs.~\cite{bs,lmr} for discussions on the 
current status of the formalisms.)

A notable characteristic in these new developments is that 
the conventional Bethe-Salpeter equation is no longer being used
to calculate the spectrum and wave functions of boundstates. 
Instead, one starts from the non-relativistic Schr\"odinger equation
(of quantum mechanics) with the Coulomb potential.
Then one adds to the nonrelativistic 
Hamiltonian relativistic corrections and radiative corrections as
perturbations to obtain an effective Hamiltonian 
(quantum mechanical operator) valid up to a necessary
order of perturbative expansion.
Effective Hamiltonians used in these new formalisms are known to be
dependent on the choice of gauge.

The purpose of this paper is to investigate gauge dependence of a
nonrelativistic boundstate formalism used in contemporary calculations
\cite{labelle},\cite{cmy},\cite{py}-\cite{yakovlev},\cite{momdist}, 
which is also closely tied to 
the potential-NRQCD formalism \cite{pNRQCD}. 
Our motivations for the study are:
(I)
In the present frontier calculations of higher order corrections
to physical quantities of boundstates, often the Feynman gauge is
used to calculate typically ultraviolet radiative corrections
whereas the Coulomb gauge is used to calculate corrections
originating typically from infrared regions.
Although much care has been taken to calculate consistently in each gauge 
separately gauge independent subsets of the corrections, 
it is desirable to clarify gauge dependences
of the formalisms actually used in these calculations.
(II)
We would like to find transformations of boundstate wave functions
when we change the gauge-fixing condition.
We may apply these transformations to study various amplitudes involving
boundstates.
Since a physical amplitude is gauge independent, once
we know how the wave function transforms, we know how
other parts of the amplitude should transform to cancel gauge
dependence as a whole.
This would provide a useful cross check for identifying all 
the contributions that have to be taken into account at a 
given order of perturbative expansion.

Already some time ago, gauge independence of the mass spectrum
of the NRQED boundstates was shown and studied in detail
based on the Bethe-Salpeter formalism \cite{by}-\cite{ffh}:
Ref.~\cite{by} gave a brief discussion;
Ref.~\cite{love} examined a Feynman gauge calculation of the 
boundstate spectrum at next-to-leading order very closely and
showed that an infinite number of two-particle irreducible diagrams
contribute in this gauge, which in the end all cancel 
due to fairly complicated gauge cancellations 
(This feature is much more complicated
in comparison to the calculation in Coulomb gauge.);
Refs.~\cite{ffh} gave formal arguments as well as perturbative analyses
which apply to all orders of perturbation series.

In comparison to these earlier works, 
new achievements of the present work are:
\begin{itemize}
\item
We use the BRST symmetry to formulate our arguments, which
allows us to treat both the NRQED and NRQCD boundstates on an equal
footing.
In particular we are able to discuss gauge dependence
of the NRQCD boundstate formalism 
rigorously using this symmetry.
\item
Presently, there exist several different definitions of an
effective Hamiltonian beyond leading order.
We introduce an effective Hamiltonian defined naturally
in the context of time-ordered (or ``old-fashioned'') perturbation 
theory of QED/QCD.
Then we obtain a transformation charge $Q$ (quantum mechanical operator)
such that the effective Hamiltonian and the boundstate wave function 
change as
\bea
&&
\delta H_{\rm eff}(P^0) = 
\left[ H_{\rm eff}(P^0) - P^0 \right] \, i Q (P^0)
- i Q^\dagger (P^0) \, \left[ H_{\rm eff}(P^0) - P^0 \right]  ,
\nonumber \\
&&
\delta \varphi = - i \,Q \cdot \varphi
\nonumber
\eea
when the gauge-fixing condition is varied infinitesimally.\footnote{
One may conjecture that gauge dependence can be described by
unitary transformation, since effective
Hamiltonians are hermite (if we neglect
decay widths of boundstates) and the boundstate spectrum is invariant
under this transformation.
For our effective Hamiltonians, however,
the transformation is not unitary (the charge is not hermite), since
the Hamiltonians are dependent on the c.m.\ energy $P^0$
of the system. 
}
Also, gauge independence of the spectrum is shown using the
transformation.
We define the charge $Q$ of the effective Hamiltonian
directly in terms of the BRST charge and the field operators in
the QED/QCD Lagrangian.

\item
For illustration: (1)we calculate the transformation charge $Q$ at
next-to-leading order;
(2)we demonstrate gauge cancellations among diagrams 
by examining an infinitesimal gauge transformation of
the amplitude for a $q\bar{q}$ boundstate decaying into 
$q'\bar{q}''W^+ W^-$.
From the latter example, one can deduce that the present calculations of
the top momentum distribution in the $t\bar{t}$ threshold region
at next-to-next-to-leading order
are gauge dependent.

\end{itemize}

Another subject of this paper is to study 
the problems in a determination of the effective
Hamiltonian from the on-shell scattering amplitude of 
a fermion and an antifermion.
Generally a fermion and an antifermion inside a boundstate are
off-shell so that use of a Hamiltonian determined in the
on-shell matching procedure may lead to incorrect calculations
of the physical quantities of the boundstate.
We clarify this point.

The same problems do not occur if we use a local
NRQED/NRQCD Lagrangian and determine 
(Wilson) coefficients in the Lagrangian
by matching onto the full theory, i.e.\ by
matching on-shell amplitudes of all relevant physical processes 
to those of perturbative QED/QCD in nonrelativistic regions.
In this case one should calculate amplitudes for
a number of processes to determine
all the coefficients.
The problems are also related to the use of the equation of motion
to simplify the Lagrangian since the on-shell condition is
the equation of motion for an asymptotic field.
For comprehensiveness we prove in an appendix that it is justified to
use the equation of motion to simplify the 
local NRQED/NRQCD Lagrangian and also
to simplify local current operators;
to the best of our knowledge such a proof has never been provided
explicitly, although similar proofs for other effective field theories
have been given \cite{lw,sf} and the claim itself is widely accepted 
already.

Below we will use the language of QCD consistently;
nevertheless all
of our arguments hold also for the QED boundstates.
Throughout the paper we 
neglect non-perturbative effects (those effects which
are typically parametrized
by $\Lambda_{\rm QCD}$) and restrict ourselves 
to arguments which can be understood from a summation of perturbation
series in $\alpha_S$ to all orders.
\medbreak

The organization of this paper is as follow.
After reviewing general aspects of gauge dependence of
the conventional relativistic $q\bar{q}$ boundstate formalism
(Sec.~\ref{s2}), we summarize characteristic 
properties of
the nonrelativistic boundstates from the viewpoint of the
leading Coulomb singularities: their gauge independence and
some nontrivial features are explained (Sec.~\ref{s3}).
Then we define the effective Hamiltonian
$H_{\rm eff}$ for a $q\bar{q}$ system and 
investigate its gauge dependence as well as gauge dependences of
the spectrum and wave functions of the boundstates using the
BRST symmetry, within the framework of perturbative expansions
in $1/c$;
in particular we define the transformation charge $Q$ of the effective
Hamiltonian (Sec.~\ref{s5}, App.~\ref{appb}).
We clarify possible problems in the determination of
$H_{\rm eff}$ if one uses the on-shell matching procedure
(Sec.~\ref{s6}).
For illustration, we present a calculation of the
charge $Q$ at ${\cal O}(1/c)$, corresponding to
an infinitesimal gauge transformation from the Coulomb gauge;
we also examine gauge cancellations in the decay amplitude of a 
$q\bar{q}$ boundstate into $q'\bar{q}''W^+W^-$ at
the same order (Sec.~\ref{s7}).
Conclusion and discussion are given in Sec.~\ref{s8}.
In App.~\ref{appa} we give a proof to justify the use of 
the equation of motion to
simplify a local NRQCD Lagrangian.
Some detailed discussions are given in App.~\ref{app2} and \ref{appb}.

\section{Gauge Dependence of the Relativistic Boundstate Formalism: 
General Aspects}
\label{s2}
\clfn

We consider the BRST invariant QCD Lagrangian
\bea
{\cal L} = -\frac{1}{2} \, {\rm tr} \left[ G^{\mu \nu} G_{\mu \nu} \right]
+ \sum_f \bar{\psi}_f [ i \! \not \! \! D(A) - m_f ] \, \psi_f +
{\cal L}_{\rm GF+FP} ,
\eea
where generally the sum of gauge-fixing and ghost terms can
be written in a BRST exact form as
\bea
{\cal L}_{\rm GF+FP} = \left\{ i Q_B, {\rm tr} [ \bar{c}F ] \right\} 
\eea
with the BRST charge $Q_B$ and an arbitrary gauge-fixing 
function $F=F(A,\psi,\bar{\psi},c,\bar{c},B)$ 
\cite{canonical}.\footnote{
In our convention, the BRST transformation is defined as
$\{ i Q_B , \psi \} = \delta_B \psi = i g \, c \, \psi$, 
$\delta_B \overline{\psi} = ig \overline{\psi} c$,
$\delta_B A_\mu = D_\mu c$,
$\delta_B c = i g \, c^2$,
$\delta_B \bar{c} = i B$,
and
$\delta_B B = 0$,
where $B$ is the Nakanishi-Laudrup auxiliary field.
}

\begin{figure}[tbp]
  \begin{center}
    \includegraphics[width=3cm]{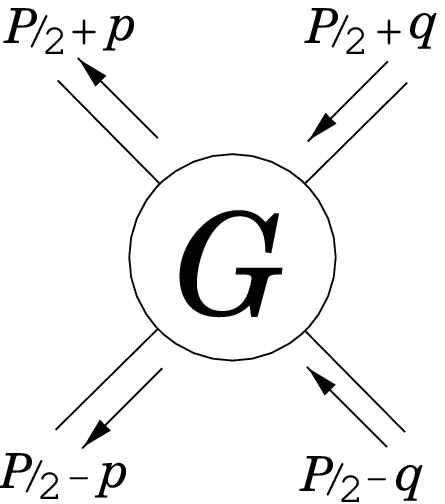}\\
    \vspace{8mm}
    \caption{\small \label{covgreen}
The four-point function $G(pqP)$.  
$P/2 \pm q$ denotes the four-momentum of incoming quark/antiquark;
$P/2 \pm p$ denotes the four-momentum of outgoing quark/antiquark.
    }
  \end{center}
\end{figure}

Define a four-point function (Fig.~\ref{covgreen}) as
\bea
G(x_1,x_2,x_3,x_4) &=& \bra{0} T \, 
\psi (x_1) \overline{\psi}(x_2) 
\overline{\psi}(x_3) \psi (x_4) 
\ket{0}
\\
&=&
\int 
{\textstyle 
\frac{d^4p}{(2\pi )^4}
\frac{d^4q}{(2\pi )^4}
\frac{d^4P}{(2\pi )^4}
}
\, G(p \, qP) 
\nonumber \\ &&
\times \exp \biggl[
-i {\textstyle \frac{P}{2}\cdot (x_1+x_2-x_3-x_4) - i p \cdot (x_1-x_2)
+ iq \cdot (x_3-x_4)} \biggr] .
\label{4ptfn}
\eea
Here and hereafter
we restrict our discussions to a quark/antiquark of some
specific flavor and omit the flavor index $f$.
The field operators and states are those of the 
Heisenberg picture.
A quark-antiquark boundstate contributes a pole 
to the Green function $G(p \, qP)$.
In the vicinity of a pole corresponding to a
boundstate $\ket{\nu ; \vec{P}}$
(with quantum number $\nu$, mass $M_\nu$ and 
momentum $\vec{P}$), the Green function
takes a form \cite{lmt}
\bea
G(p \, qP) = \frac{i}{2\omega_{\nu ,\vec{P}}} \,
\frac{\chi_{\nu,\vec{P}}(p)\overline{\chi}_{\nu,\vec{P}}(q)}
{P^0 - \omega_{\nu ,\vec{P}} + i \epsilon }
\, + (\mbox{regular as }P^0 \to \omega_{\nu ,\vec{P}} ) ,
\label{pole}
\eea
where
\bea
&&
\bra{0}T \, \psi (x_1) \overline{\psi}(x_2) \ket{\nu;\vec{P}}
= e^{-i(\omega_{\nu ,\vec{P}} X^0 - \vec{P}\cdot \vec{X})}
\int {\textstyle \frac{d^4p}{(2\pi )^4} } 
\, e^{-ip\cdot x} \, \chi_{\nu,\vec{P}}(p) ,
\\ &&
\overline{\chi}_{\nu,\vec{P}}(p)=
\chi_{\nu,\vec{P}}(p)^\dagger (\gamma^0 \otimes \gamma^0) ,
\\ &&
{\textstyle X = \frac{1}{2}(x_1+x_2), ~~~ x=x_1-x_2} , ~~~
\omega_{\nu ,\vec{P}} = \sqrt{\vec{P}^2+M_\nu^2} .
\eea
In this paper we assume that the decay width of a boundstate is
infinitesimally small except where it is stated otherwise.

An infinitesimal deformation of the gauge-fixing function,
$F \to F + \delta F$, induces a change of the Lagrangian
\bea
\int d^4x \, \delta {\cal L} = \{ i Q_B , \delta {\cal O} \} ,
~~~~~ ~~~~~
\delta {\cal O} \equiv \int d^4x \, {\rm tr} [ \bar{c} \, \delta F ] .
\eea
Accordingly the Green function changes as
\bea
\delta G(x_1,x_2,x_3,x_4) &=&
- \bra{0} T \, \{ Q_B , \delta {\cal O} \}  \,
\psi (x_1) \overline{\psi}(x_2) 
\overline{\psi}(x_3) \psi (x_4) 
\ket{0}
\\ 
&=& - \bra{0} T \, \delta {\cal O} \, [ Q_B ,
\psi (x_1) \overline{\psi}(x_2) 
\overline{\psi}(x_3) \psi (x_4) ]
\ket{0} .
\eea

Suppose we are interested in the boundstates which can be created 
from the vacuum via 
gauge invariant operators,
e.g. $\overline{\psi}(x) \psi (x),~
\overline{\psi}(x) \! \not{\! \! D} \psi (x),~
\overline{\psi}(x) \gamma^\mu \psi (x)$, etc.
For example, in the above Green function
we may set $x_1=x_2 \equiv x$, $x_3=x_4 \equiv y$ and contract 
color indices to make color singlet operators 
$\overline{\psi} (x) \psi (x)$, $\overline{\psi}(y) \psi (y)$.
It then follows from $[Q_B, \overline{\psi} \psi ]=0$
that
\bea
\delta G(x,x,y,y) \biggr|_{\mbox{\scriptsize color singlet}} = 0 .
\eea
Comparing this with eq.~(\ref{pole}), one sees that both the mass and the
residue of any boundstate which couples to the operator 
$\overline{\psi} \psi$ are invariant:
\bea
\delta M_\nu = 0
\eea
and
\bea
\delta \bra{0} \, \overline{\psi}(x)  \psi (x) \ket{\nu;\vec{P}} = 0 ,
~~~~~ ~~~~~
\mbox{i.e.}
~~
\delta \int {\textstyle \frac{d^4p}{(2\pi )^4} } 
\, \chi_{\nu,\vec{P}}(p) 
\biggr|_{\mbox{\scriptsize color singlet}} = 0 .
\eea
Since both $\ket{0}$ and $\overline{\psi}(x) \psi (x)$ are
BRST invariant, it implies that the boundstate satisfies the
physical state condition
\bea
Q_B \ket{\nu;\vec{P}} = 0 .
\label{physstatecond}
\eea
Note, however, that generally the boundstate wave function
$\chi_{\nu,\vec{P}}(p)$ depends
on the gauge-fixing condition:
\bea
\delta \chi_{\nu,\vec{P}}(p) &=& - \, \mbox{F.T.} \,
\bra{0}T \, \delta {\cal O} \, [ Q_B, \psi (x) \overline{\psi}(y) ]
\ket{\nu;\vec{P}}
\nonumber \\
&=& - \, \mbox{F.T.} \, 
\bra{0}T \, g \, \delta {\cal O} \,
[ c(x)\psi(x) \overline{\psi}(y) + 
\psi (x) \overline{\psi}(y) c(y) ]
\ket{\nu;\vec{P}} ,
\label{delchi}
\eea
where F.T. stands for an appropriate Fourier transform.

Any physical amplitude $\braket{f;{\rm out}}{i;{\rm in}}$ which involves
the quark-antiquark boundstate contributions includes the above Green 
function $G(pqP)$ as a part of it.
Since the initial and final states satisfy the physical state conditions
$Q_B \ket{i;{\rm in}} = Q_B \ket{f;{\rm in}} = 0$ and the theory
is BRST invariant, the amplitude is gauge independent.
Hence, the boundstate poles included in the amplitude 
are also gauge independent.
An interesting question is whether the Green function $G(pqP)$ 
includes any unphysical pole, which does
not contribute to the physical amplitude, close to or degenerate
with one of the physical boundstate poles.\footnote{
A typical example is the $R_\xi$-gauge for electroweak
interaction where an unphysical pole $(k^2-\xi M_W^2 + i\epsilon )^{-1}$
is included in the gauge boson propagator.
}
As for nonrelativistic quark-antiquark boundstates the
answer is no,
as will be shown in Section~\ref{s5} and in Appendix~\ref{appb}.

\section{Nonrelativistic Boundstates: Leading Coulomb Singularities}
\label{s3}
\clfn

It is well-known that, in describing a system
of a nonrelativistic color-singlet quark-antiquark ($q\bar{q}$) pair,
naive perturbation theory breaks down 
due to formation of boundstates 
\cite{braun,ap}.
Intuitively, this is because the slow $q$ and $\bar{q}$ 
are trapped by the attractive force mediated by exchange of
gluons and multiple exchange of gluons between the pair 
becomes significant.
We review this property 
in a production process of a $q\bar{q}$ pair.

\begin{figure}[tbp]
  \begin{center}
    \includegraphics[width=13cm]{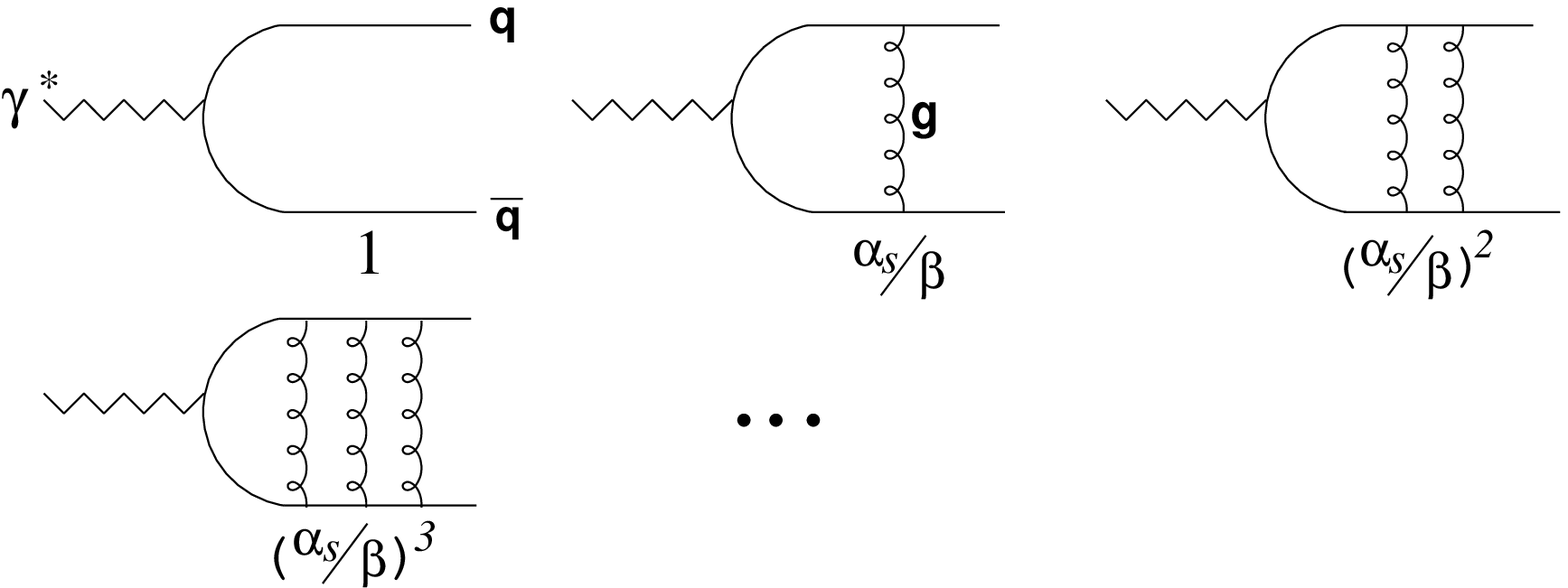}\\
    \vspace{8mm}
    \caption{\small \label{fig1}
      The ladder diagrams for the process 
      $\gamma^* \to q\bar{q}$.
      The diagram where $n$ uncrossed gluons are exchanged has a behavior
      $\sim (\alpha_S/\beta)^n$ near threshold.
    }
  \end{center}
\end{figure}

Consider the amplitude of a virtual photon 
decaying into $q$ and $\bar{q}$,
$\gamma^* \to q\bar{q}$, just above the threshold.
As we will see below, the ladder diagram for this
process with $n$ gluon exchanges 
has a behavior $\sim (\alpha_S/\beta)^n$, see Fig.~\ref{fig1}.
Here, $\beta$ is
the velocity of $q$ or $\bar{q}$ in the c.m.\ frame,
\bea
\beta = \sqrt{1-\frac{4m^2}{s}} .
\label{beta}
\eea
Hence, the contribution of the $n$-th ladder diagram
will not be small even for a large $n$ if $\beta \simlt \alpha_S$.
That is, higher order terms in $\alpha_S$ remain unsuppressed
in the threshold region.
These singularities in $\beta$ which appear in this specific kinematical
configuration is known as the ``Coulomb singularities'' or
``threshold singularities''.
The singularities arise because, for a particular
assignment of the loop momenta, all the internal particles can simultaneously 
become almost on-shell as $\beta \to 0$.

\begin{figure}[tbp]
  \begin{center}
    \includegraphics[width=14cm]{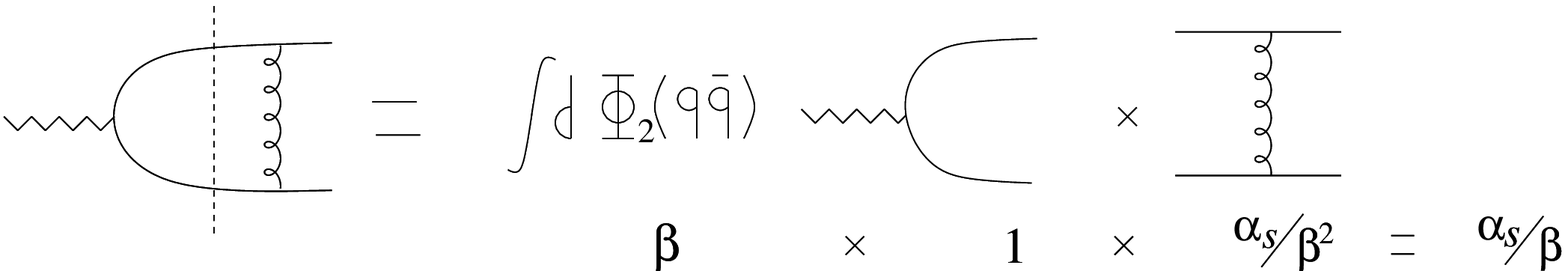}\\
    \vspace{8mm}
    \caption{\small \label{fig2}
      The Cutkosky rule for evaluating the imaginary part of the 1-loop
      diagram.  The factors in $\alpha_S$ and $\beta$ are shown explicitly.
    }
  \end{center}
\end{figure}

\begin{figure}[tbp]
  \begin{center}
    \includegraphics[width=15cm]{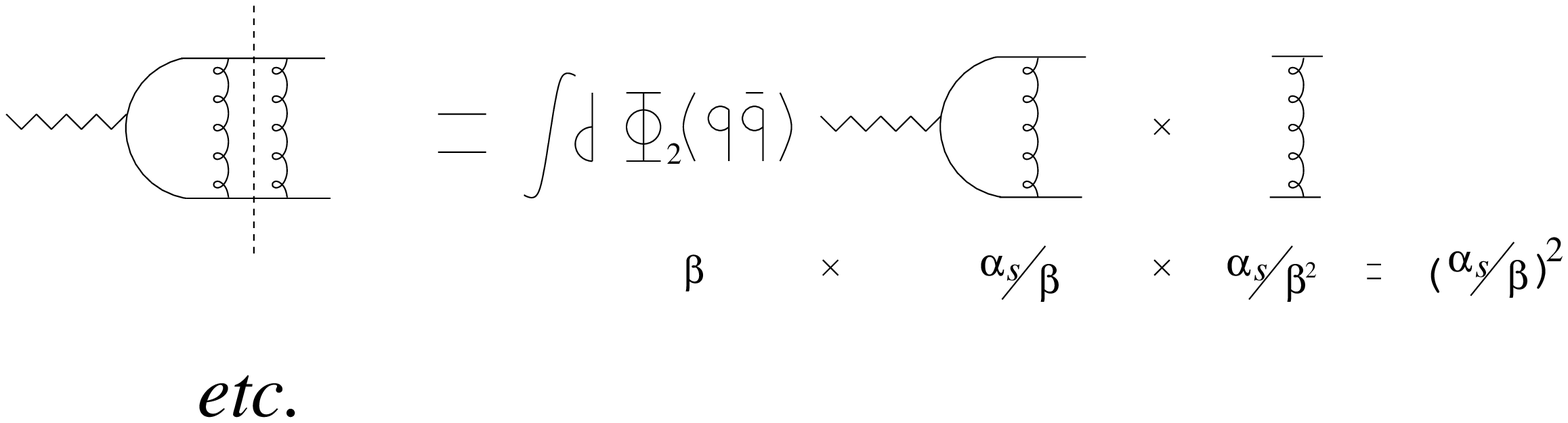}\\
    \vspace{8mm}
    \caption{\small \label{fig3}
      The cut-diagram method for evaluating the singularities of the
      higher order ladder diagrams.  The factors in $\alpha_S$ and $\beta$ are
      shown explicitly.
    }
  \end{center}
\end{figure}

The appearance of the factor $(\alpha_S/\beta )^n$
may be seen as follows.
First, consider the one-loop diagram.
Its {\it imaginary} part can be estimated using the Cutkosky rule
(cut-diagram method), see Fig.~\ref{fig2}.
Namely, the imaginary part is
given by the phase space integration of the product of 
tree diagrams.
The intermediate $q\bar{q}$ phase space is proportional to $\beta$ as
\bea
d\Phi_2(q\bar{q}) = \frac{\beta}{16\pi} \, d \cos \theta ,
\eea
where $\theta$ is the angle between the momenta of 
the intermediate and final quarks in the c.m.\ frame.
The $q\bar{q}$ scattering diagram with 
a $t$-channel gluon exchange
contributes a factor $\sim \alpha_S/\beta^2$ since the
gluon propagator is proportional to $1/\beta^2$;
the propagator denominator is given by
\bea
k^2 = 
-|\vec{k}|^2 = - \, \frac{s \beta^2}{2} (1-\cos \theta),
\label{glmomsq}
\eea
where $k$ denotes the gluon momentum.
Thus, we see that the imaginary part of the one-loop diagram
has the behavior $\sim \beta \cdot \alpha_S /\beta^2 = \alpha_S/\beta$.
Analyticity implies that the real part of the one-loop diagram
has the same structure $\sim \alpha_S/\beta$.
By repeatedly using the cut-diagram method, one can show by induction
that the imaginary part of the 
ladder diagram with $n$ uncrossed gluons behaves as
$\sim (\alpha_S/\beta)^n$, see Fig.~\ref{fig3}.

Alternatively this fact can be shown by a power counting method \cite{bs}.
The relevant loop momenta in the loop integrals are 
in the nonrelativistic regime:
\bea
p^0 - m, \, \bar{p}^0-m \sim {\cal O}(\beta^2),
&&~~
\vec{p} = -\vec{\bar{p}} \sim {\cal O}(\beta),
\nonumber
\\
k^0 \sim {\cal O}(\beta^2), ~~
&&
\vec{k} \sim {\cal O}(\beta) .
\label{nrmom}
\eea
Here, $p$, $\bar{p}$ and $k$ represent the internal momenta of
$q$, $\bar{q}$ and the gluon, respectively, in the c.m.\ frame.
For such configurations, $q(\bar{q})$ and gluon
propagators are counted as $\sim 1/\beta^2$, and the
measure for each loop integration $d^4k/(2\pi)^4$ as
$\sim \beta^5$.\footnote{
In counting the powers of $\beta$ of a loop integral,
the singularity of the integrand will increase if one
assigns a large power of $\beta$ to the momentum in the propagators, 
but the integration measure is more suppressed.
The optimal assignment of the order in $\beta$ to 
each internal momentum must be sought to identify the most singular 
part of the integral.
This procedure leads to Eq. (\ref{nrmom}).
}

Thus, the ladder diagrams contain the leading singularities
$\sim (\alpha_S/\beta)^n$.
Other diagrams (in particular crossed gluon diagrams) do not possess 
the leading singularities but only non-leading
singularities $\sim \alpha_S^{n+l}/\beta^n~$ ($l \geq 1$).

As the higher order terms in $\alpha_S$ cannot be neglected
near threshold, we are led to sum up the leading Coulomb
singularities.
Let us first discuss gauge dependence of the amplitude when
this summation is performed, in particular because
only the ladder diagrams are included.
The exact amplitude for the process
$e^+e^- \to q\bar{q}$ near threshold can be expanded in terms of $\alpha_S$ and
$\beta$ as
\bea
{\cal M}^{(full)} (\alpha_S,\beta)
= \sum_{n=0}^\infty \, c_n (\alpha_S/\beta)^n
\, + \, (\mbox{non-leading terms}).
\eea
The full amplitude ${\cal M}^{(full)}$ is gauge independent, so must be
the each coefficient $c_n$.
Because only the ladder diagrams possess this type of singularities, 
{\it the most singular part} of the ladder diagrams has to be gauge
independent.

To see this explicitly, we examine gauge dependence of the gluon 
propagator.
In a general covariant gauge, 
the $q\bar{q} \to q\bar{q}$ scattering amplitude 
in the threshold region is given by 
\bea
&&
\bar{u}_f \gamma^\mu u_i \, \,
\frac{-i}{k^2+i\epsilon} 
\left[
g_{\mu \nu} - ( 1 \! - \! \xi ) \frac{k_\mu k_\nu}{k^2}
\right]
\bar{v}_i \gamma^\nu v_f 
=
\bar{u}_f \gamma^0 u_i \, \,
\frac{ -i }{-\vec{k}^2 + i \epsilon} 
\, \,
\bar{v}_i \gamma^0 v_f ~\times ~
\biggl[ 1 + {\cal O}(\beta) \biggr] ,
\nonumber \\
\label{glp}
\eea
where the subscripts $i$ and $f$ stand for the initial and
final state, respectively.
We have used the fact that the space components of the currents,
$\bar{u}_f \gamma^\mu u_i$ and $\bar{v}_i \gamma^\nu v_f$,
are order $\beta$ in the c.m.\ frame.\footnote{
Dirac representation of the $\gamma$-matrices is most useful
in power countings, where $\gamma^0$ is diagonal and
$\gamma^i$'s are off-diagonal.
The quark
spinor wave function has the upper two components of 
${\cal O}(1)$ and the lower two components suppressed by $\beta$,
{\it vice versa} for the antiquark.
}
Note that the leading part of the gluon propagator 
is identical with the Coulomb propagator in the Coulomb gauge.
Eq.~(\ref{glp}) also holds for the momenta (\ref{nrmom}) 
if we note that the 
off-shell $q$ and $\bar{q}$ wave functions are given
by
\bea
\not \! p + m = m (1+\gamma^0) + {\cal O}(\beta) ,
\\
- \! \not \! \bar{p} + m  = m (1-\gamma^0)
+ {\cal O}(\beta) .
\eea
Thus, gauge independence of $c_n$'s is ensured
by gauge independence of the leading part $\sim 1/\beta^2$
of the gluon propagator in eq.~(\ref{glp}).

\begin{figure}[tbp]
  \begin{center}
    \includegraphics[width=13cm]{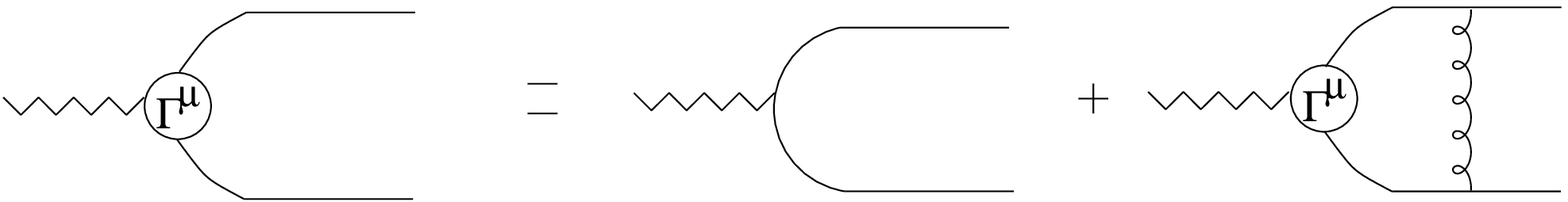}\\
    \vspace{8mm}
    \caption{\small \label{fig4}
      The self-consistent equation satisfied by the leading singularities
      of the $q\bar{q}\gamma$ vertex $\Gamma^\mu$.  One should take only the
      leading part $\sim (\alpha_S/\beta)^n$ on both sides of the equation.
    }
  \end{center}
\end{figure}

Let us denote by $\Gamma^\mu$ the sum of the leading singularities of the
vertex $\gamma^* \to q\bar{q}$.
It satisfies a self-consistent
equation as depicted in Fig.~\ref{fig4}.
Retaining only the leading part $\sim (\alpha_S/\beta)^n$ on
both sides of the equation, one obtains the
vertex $\Gamma^\mu$ as
\bea
\Gamma^\mu = - \,
\left( \frac{1+\gamma^0}{2} \gamma^\mu \frac{1-\gamma^0}{2} \right)
\, ( E - \vec{p}\, ^2/m  ) \,
\tilde{G}( \vec{p};E ),
\eea
where $E = \sqrt{s}-2m$ is the energy measured from the threshold.
$\tilde{G}( \vec{p};E )$ is the Green function
of the nonrelativistic Schr\"{o}dinger equation with the
Coulomb potential:
\bea
&&
\left[
\left( - \, \frac{\nabla^2}{m} - \, C_F \, \frac{\alpha_S}{r} \right) 
- ( E + i\epsilon )
\right]
G (\vec{r};E) = \delta^3(\vec{r}),
\\ && ~~~
\tilde{G}( \vec{p};E )
= \int d^3 \vec{r} \, e^{-i\vec{p} \cdot \vec{r} }
\, G (\vec{r};E) ,
\eea
where $C_F = 4/3$ is the color factor.
The analytic expression of $G(\vec{r};E)$ is 
given in terms of the hypergeometric function and
includes the boundstate spectrum below threshold, $E<0$.
Alternatively, we may write
\bea
\tilde{G}(\vec{p};E) = - 
\hbox{ \hbox to 0pt{${\displaystyle \sum_n}$}\hbox{
${\displaystyle \int}$} }
\frac{\phi_n (\vec{p})\psi_n^* (\vec{0})}{E-E_n+i\epsilon},
\label{swgf0}
\eea
where $\phi_n (\vec{p})$ and $\psi_n (\vec{r})$ are the
Coulomb wave functions in momentum space and coordinate space,
respectively.
Here, $n$ includes the boundstates with $E_n=-(C_F\alpha_S)^2 m/4n^2$
and the continuum states with $E_n > 0$.\footnote{
To see that
$(E-\vec{p}\, ^2/m)\tilde{G}(\vec{p};E)$ is a function of $\alpha_S/\beta $,
one should identify $E \to m \beta^2$ and $|\vec{p}| \to m \beta$
at leading order.
}

At this stage, 
we see a nontrivial consequence of the summation to all orders in
$\alpha_S$.
At any order of the perturbative expansion in $\alpha_S$,
the amplitude for $\gamma^* \to q \bar{q}$ is zero below threshold,
$E<0$.
For example, the absorptive part of a
quark loop contribution to the vacuum polarization function
\bea
( P^2 g^{\mu \nu} - P^\mu P^\nu ) \, {\rm Im} \, \Pi_q (P^2) 
= 
\int d^4x \, e^{i P \cdot (x-y)} \,
{\rm Im} \, \bra{0} T \, 
\overline{\psi} (x) \gamma^\mu {\psi}(x) 
\overline{\psi}(y) \gamma^\nu \psi (y) 
\ket{0}
\eea
vanishes below threshold.
After summation of the leading singularities, however, it is
given in terms of the Green function at the origin \cite{braun}
\bea
{\rm Im} \, \Pi_q (s) = 
\frac{N_c}{2m^2} \, {\rm Im} \, G (\vec{r}=0;E=\sqrt{s}-2m) 
= \frac{\pi N_c}{2m^2} \, 
\hbox{ \hbox to 0pt{${\displaystyle \sum_n}$}\hbox{
${\displaystyle \int}$} }
|\psi_n(\vec{0})|^2 \, \delta ( E - E_n ) ,
\label{impi}
\eea 
which 
in fact diverges at the positions of boundstates, $E= E_n<0$.
This discrepancy before and after the summation can be
traced back to the fact that
the limit $\epsilon \to 0$ in the propagator denominators does not commute
with the summation to infinite orders in $\alpha_S$.
Namely, if we pursue the perturbative calculations with a 
finite $\epsilon > 0$,
the absorptive part ${\rm Im} \, \Pi_q(s)$ remains nonzero below threshold at
each order.
After the summation, constructive interference effects
result in a drastic
magnification of the amplitude $\sim 1/\epsilon$ at $E= E_n$.

In order to reach below the threshold for the process 
$e^+e^- \to q\bar{q}$, we need to include a subsequent decay process,
e.g. $q$ and $\bar{q}$ decaying into lighter quarks, or $q\bar{q}$ 
annihilating into multiple gluons, etc.
Then the corresponding
amplitude is nonzero and gauge independent both above and
below the threshold.
Summation of the leading singularities can be performed in the
same way as above and leads to the same vertex $\Gamma^\mu$,
except that in this case
the quark momentum $|\vec{p}|$ needs not equal $\sqrt{s/4-m^2}$
(as required for an on-shell quark) as long as
it is in the nonrelativistic region.\footnote{
The nonzero decay width of the boundstate $\Gamma_n$
renders the $\delta$-function in eq.~(\ref{impi})
to the Breit-Wigner distribution
$$
\pi \delta (E-E_n) \to \frac{\Gamma_n/2}{(E-E_n)^2+\Gamma_n^2/4} .
$$
}

Below we discuss gauge dependences of the
spectrum and the wave functions of the nonrelativistic $q\bar{q}$
boundstates 
within the framework of their calculations in perturbative expansions.
An appropriate expansion parameter of this problem is $1/c$,
inverse of the speed of light, when $c$ is restored as a
dimensionful parameter \cite{gr}.
In this case both $\alpha_S = g^2/4\pi\hbar c$ and 
$\beta = v/c$ are ${\cal O}(1/c)$ quantities.\footnote{
Here, $\beta$ symbolizes both $|\vec{p}|/mc$ and 
$\sqrt{E/mc^2}$ for a nonrelativistic off-shell quark/antiquark.
}
Therefore the sum of the leading singularities
$(\alpha_S/\beta)^n$ is counted as ${\cal O}(1)$.
Perturbative corrections to the boundstate wave function are 
given as a double expansion in $\alpha_S$ and $\beta$,
e.g. ${\cal O}(1/c)$ corrections include
$\alpha_S^{n+1}/\beta^n = \alpha_S (\alpha_S/\beta)^n
= \beta (\alpha_S/\beta)^{n+1}$.
Note that the parameter $\beta$ is guaranteed to be
small if $\alpha_S$ is small, since we are interested in the
summation of the leading singularities only in the kinematical
region where naive perturbation theory breaks down 
$(\beta \simlt \alpha_S)$.
The boundstate mass is given as a power series in $\alpha_S$ since
they are independent of $\beta$.\footnote{
In addition to powers of $\alpha_S$ and $\beta$, there appear
also powers of $\log \alpha_S$ and $\log \beta$ in these perturbation
series.
}
Throughout this paper we set $c=1$ in our formulas
in order to maintain simplicity of expressions;
one may easily count the power of $1/c$ by counting the powers of 
$\alpha_S$ and $\beta$.

\section{The Effective Hamiltonian for a $q\bar{q}$ System}
\label{s5}
\clfn

In this section we discuss gauge dependences of the
spectrum and the wave functions of the nonrelativistic 
boundstates in a general framework.

Let us introduce an effective Hamiltonian for a color-singlet
$q\bar{q}$ system as follows.
First we define a Green function for a $q\bar{q}$ pair in the c.m.\
frame as
\bea
{\cal G}(\vec{p},\vec{q};\lambda,\bar{\lambda},\lambda',\bar{\lambda}';P^0)
= \bra{\vec{p},-\vec{p},\lambda,\bar{\lambda}} 
\frac{1}{P^0-H+i\epsilon}
\ket{\vec{q},-\vec{q},\lambda',\bar{\lambda}'} ,
\label{defcalg}
\eea
where 
$H$ denotes the full QCD Hamiltonian (including the 
gauge-fixing and ghost terms);
$\ket{\vec{p},-\vec{p},\lambda,\bar{\lambda}}$
is an eigenstate of the {\it free} Hamiltonian $H_0 = H |_{\alpha_S \to 0}$
and represents
a color-singlet two-body state composed of a free quark-antiquark pair:
\bea
\ket{\vec{p},-\vec{p},\lambda,\bar{\lambda}} = 
a^\dagger_{\vec{p},\lambda} b^\dagger_{-\vec{p},\bar{\lambda}} 
\ket{0}_{\rm free} \biggr|_{\mbox{\scriptsize color singlet}}
 ,
~~~~~ ~~~~~
H_0 \ket{0}_{\rm free} = 0 .
\label{asymstate}
\eea
Here, $a^\dagger$ ($b^\dagger$) denotes the creation operator of
a free quark (antiquark); 
$\vec{p}$ ($-\vec{p}$) and $\lambda$ ($\bar{\lambda}$) denote the three
momentum and the spin index of $q$ ($\bar{q}$) in the c.m.\ frame,
respectively.
$P^0$ represents symbolically
the c.m.\ energy of the $q\bar{q}$ system, but we take the three
energies, $P^0$, $2\sqrt{\vec{p}\, ^2+m^2}$ and 
$2\sqrt{\vec{q}\, ^2+m^2}$,
not necessarily equal to one another. 
Note that the above two-body state is not a physical state,
$Q_B \ket{\vec{p},-\vec{p},\lambda,\bar{\lambda}} \neq 0$,
which stems from the fact that $H_0$ is not BRST invariant.
Then we define an effective Hamiltonian which operates only on the
subspace spanned by the two-body states
such that it generates the same Green function:
\bea
{\cal G}(\vec{p},\vec{q};\lambda,\bar{\lambda},\lambda',\bar{\lambda}';P^0)
= \bra{\vec{p},-\vec{p},\lambda,\bar{\lambda}} 
\frac{1}{P^0-H_{\rm eff}(P^0)+i\epsilon}
\ket{\vec{q},-\vec{q},\lambda',\bar{\lambda}'} .
\eea
Namely, the effective Hamiltonian (a quantum mechanical
operator) is defined by
\bea
H_{\rm eff}(P^0)
= P^0 
- {\cal G}^{-1} (P^0) ,
\label{defheff}
\eea
where 
${\cal G}^{-1}
(\vec{p},\vec{q};\lambda,\bar{\lambda},\lambda',\bar{\lambda}';P^0)
= \bra{\vec{p},-\vec{p},\lambda,\bar{\lambda}} {\cal G}^{-1} (P^0)
\ket{\vec{q},-\vec{q},\lambda',\bar{\lambda}'}$
is the inverse of the Green function restricted to the two-body
subspace [take the inverse of
${\cal G}(\vec{p},\vec{q};\lambda,\bar{\lambda},\lambda',\bar{\lambda}';P^0)$
considering it
to be a matrix with indices $(\vec{p},\lambda,\bar{\lambda})$ 
and
$(\vec{q},\lambda',\bar{\lambda}')$].
For analyzing the nonrelativistic $q\bar{q}$ boundstates,
one first calculates the effective
Hamiltonian in a series expansion in $1/c$,
then uses ordinary perturbation theory in quantum mechanics
for calculating the spectrum and the wave functions of the boundstates
in perturbative expansions in $1/c$.
As we have seen in the previous section, the leading order Hamiltonian is
given by\footnote{
Presently the QCD effective Hamiltonian is known up to
${\cal O}(1/c^2)$ in Coulomb gauge, see e.g.\ \cite{py,NNLO}.
}
\bea
H_{\rm eff}^{\rm (LO)} = 2 m + 
\frac{\vec{p}\, ^2}{m} - C_F \frac{\alpha_S}{r}  .
\eea

Let us briefly explain the background why we introduced the Green function,
eq.~(\ref{defcalg}).
Suppose we consider
contributions from a $q\bar{q}$ boundstate to some physical process.
In a calculation of the corresponding amplitude using time-ordered
(or ``old-fashioned'') perturbation theory,
the above Green function always appears as a part of that calculation.
This is parallel to the fact that the four-point function
eq.~(\ref{4ptfn}) appears as a part of the calculation of
the same amplitude using the (Lorentz covariant) Feynman rules.
Time-ordered perturbation theory is often more suited for calculations of
nonrelativistic processes because additional quark-antiquark pair productions
are suppressed by powers of $1/c$.

The rules for time-ordered perturbation theory are \cite{text}: 
draw time-ordered
diagrams (e.g. time flows from right to left), assign a matrix element
$\bra{i} V_\alpha \ket{j}$ at
the time of each vertex, and assign a propagator $1/(P^0-E_i+i\epsilon)$
to an interval between two adjacent vertices.
Here, $V_\alpha$ is an interaction term,
${\displaystyle H = H_0 + \sum_\alpha V_\alpha}$;
$P^0$ is the total energy of the system;
$\ket{i}$ and $E_{i}$ denote the eigenstate and the eigenvalue of the
free Hamiltonian, respectively, $H_0 \ket{i} = E_i \ket{i}$.
Then we sum over all the intermediate states, where in general the
energy is not conserved, $E_i \neq P^0$.
\begin{figure}[tbp]
  \begin{center}
    \includegraphics[width=5cm]{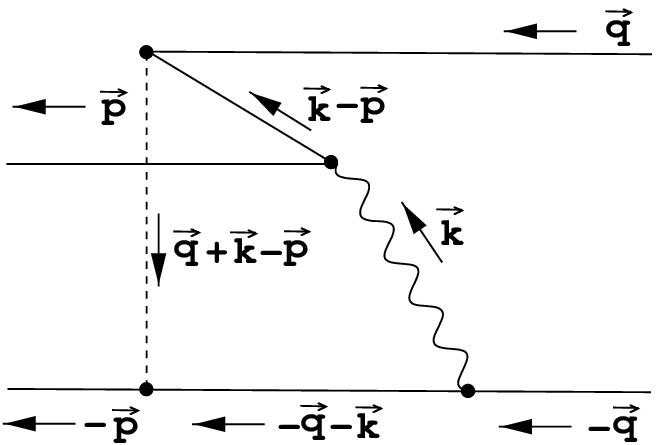}\\
    \vspace{8mm}
    \caption{\small \label{d8}
A time-ordered diagram which contributes to ${\cal G}$.
The dashed line represents the instantaneous Coulomb gluon;
the wavy line represents the transverse gluon.
    }
  \end{center}
\end{figure}

Although there are many ways to derive the rules 
(see Appendix~\ref{app2}), 
simple correspondences to the ordinary Feynman rules may be seen
by integrating over the time components of loop momenta and
over the time components 
of external particles' momenta of a Feynman diagram for an
unamputated Green function.
In Coulomb gauge,
decomposing the quark and transverse gluon propagators as
\bea
&&
\frac{i ( \not \! p + m )}{p^2-m^2+i\epsilon} = 
\frac{i}{p^0-\omega_{\vec{p}}+i\epsilon} \,
\, \Lambda_+ (\vec{p}) \gamma^0
+
\frac{i}{p^0+\omega_{\vec{p}}-i\epsilon} \,
\, \Lambda_- (\vec{p}) \gamma^0 ,
\label{quarkpp}
\\
&&
\omega_{\vec{p}} = \sqrt{ \vec{p} \, ^2 + m^2 } ,
~~~~~
\Lambda_\pm (\vec{p}) = 
\frac{ \omega_{\vec{p}} \pm ( m - \vec{p}\cdot \vec{\gamma} ) \gamma^0 }
{2 \omega_{\vec{p}} } ,
\\
&&
\frac{i}{k^2+i\epsilon}
\biggl( \delta_{ij} - \frac{k_i k_j}{|\vec{k}|^2} \biggr)
= \frac{i}{2|\vec{k}|}
\biggl( \delta_{ij} - \frac{k_i k_j}{|\vec{k}|^2} \biggr)
\biggl( 
\frac{1}{k^0-|\vec{k}|+i\epsilon} -
\frac{1}{k^0+|\vec{k}|-i\epsilon}
\biggr) ,
\label{tgluonpp}
\eea
and using the Cauchy theorem, every wave function becomes
on-shell 
[e.g. 
$\sum_\lambda u(\vec{p},\lambda) u^\dagger(\vec{p},\lambda)
= \Lambda_+(\vec{p})$]
whereas the energy conservation is violated.
The ghost propagator can be handled similarly to the
transverse gluon propagator.
Integrating the Coulomb propagator is trivial because it is
independent of the energy of the gluon.
By way of example, the diagram in Fig.~\ref{d8},
which contributes to the Green function ${\cal G}$, is given by
\bea
\frac{( C_F \cdot 4\pi\alpha_S )^2}
{(P^0-2\omega_{\vec{p}}+i\epsilon)
(P^0-2\omega_{\vec{q}}+i\epsilon)}
\int \frac{d^3\vec{k}}{(2\pi)^3} \, \,
{u}^\dagger (\vec{p},\lambda) \, \gamma^0 \gamma^i \,
\Lambda_-(\vec{p}-\vec{k})
\, u(\vec{q},\lambda')
\nonumber \\ 
\times
{v}^\dagger (-\vec{q},\bar{\lambda}') \, \gamma^0 \gamma^j \,
\Lambda_-(\vec{q}+\vec{k})
\, v(-\vec{p},\bar{\lambda}) \,
\frac{1}{2|\vec{k}|}
\biggl( \delta^{ij} - \frac{k^ik^j}{|\vec{k}|^2} \biggr) 
\frac{-1}{|\vec{q}+\vec{k}-\vec{p}|^2}
\nonumber \\ 
\times
\frac{1}{
(P^0-\omega_{\vec{q}}-\omega_{\vec{k}-\vec{p}}-\omega_{\vec{p}}
-\omega_{\vec{q}+\vec{k}}+i\epsilon)
(P^0-\omega_{\vec{q}}-\omega_{\vec{q}+\vec{k}}-|\vec{k}|+i\epsilon)
}
.
\eea

We return to the discussion of the Green function ${\cal G}$ and
the effective Hamiltonian $H_{\rm eff}$.
If we vary the gauge-fixing function, the QCD Hamiltonian changes
as\footnote{
Here, we assume that $\delta F$ is independent of
$\partial_0 A_\mu$, $\partial_0 c$, etc.;
otherwise the change of the Hamiltonian takes a different form.
See Section~\ref{s7} for a more general case.
}
\bea
H \to H - \{ i Q_B , \delta O \} ,
~~~~~ ~~~
\delta O = \int d^3 \vec{x} \, \, {\rm tr} [ \bar{c} \delta F ] ,
\eea
and the corresponding change of the Green function is given by
\bea
&&
\delta {\cal G}
(\vec{p},\vec{q};\lambda,\bar{\lambda},\lambda',\bar{\lambda}';P^0)
= 
- \bra{\vec{p},-\vec{p},\lambda,\bar{\lambda}} 
\frac{1}{P^0-H+i\epsilon}
\, \{ i Q_B , \delta O \} \,
\frac{1}{P^0-H+i\epsilon}
\ket{\vec{q},-\vec{q},\lambda',\bar{\lambda}'} 
\nonumber \\ &&
\\
&&
~~~~~ =
- \bra{\vec{p},-\vec{p},\lambda,\bar{\lambda}} 
i Q_B \, \frac{1}{P^0-H+i\epsilon}
\, \delta O \,
\frac{1}{P^0-H+i\epsilon}
\ket{\vec{q},-\vec{q},\lambda',\bar{\lambda}'} 
\nonumber 
\\ &&
~~~~~~~~
- \bra{\vec{p},-\vec{p},\lambda,\bar{\lambda}} 
\frac{1}{P^0-H+i\epsilon}
\, \delta O \,
\frac{1}{P^0-H+i\epsilon} \, i Q_B \,
\ket{\vec{q},-\vec{q},\lambda',\bar{\lambda}'} .
\label{deltag}
\eea
Since 
$Q_B \ket{\vec{p},-\vec{p},\lambda,\bar{\lambda}} \neq 0$,
generally $\delta {\cal G} \neq 0$, so the 
corresponding effective Hamiltonian also changes,
$\delta H_{\rm eff} \neq 0$. 
As we use the effective Hamiltonian to calculate systematically the boundstate
spectrum and the wave functions in perturbative expansions
in $1/c$, we would like 
to see how they depend on our choice of gauge.

The mass $M_\nu$ of a boundstate is given as the position of
a pole of the Green function ${\cal G}$.
Equivalently, it is calculated from $H_{\rm eff}(P^0)$
by solving
\bea
\left[ M_\nu - H_{\rm eff}(M_\nu) \right] \, \ket{\nu;{\rm eff}}
= 0.
\eea
In the discussion to follow, we consider 
only those boundstates which appear already at leading order
for $\ket{\nu;{\rm eff}}$.\footnote{
Note that at leading order
all boundstates in the spectrum are the Coulomb 
boundstates which are physical states.
(In particular, all these states
can be created from the vacuum via gauge invariant operators.)
It already suggests that to all orders of $1/c$
there are no unphysical boundstates in the
spectrum of $H_{\rm eff}$ which are degenerate
with these physical boundstates.
}
From the definition of $M_\nu$ above, one may evaluate its deviation
when the effective Hamiltonian is varied infinitesimally:
\bea
\delta M_\nu = 
\frac{
\bra{\nu;{\rm eff}} \, \delta H_{\rm eff}(M_\nu) \, \ket{\nu;{\rm eff}} 
}{
\bra{\nu;{\rm eff}} \, 
1 - H_{\rm eff}'(M_\nu) \, \ket{\nu;{\rm eff}} 
} .
\label{delmnu}
\eea
In the numerator, the variation of the Hamiltonian 
can be written as
\bea
\delta H_{\rm eff} (P^0)
&=& - \, \delta {\cal G}^{-1}(P^0) 
\nonumber \\ 
&=& 
\left[ P^0 - H_{\rm eff}(P^0) \right]
\, \delta {\cal G}(P^0) \,
\left[ P^0 - H_{\rm eff}(P^0) \right] 
\label{delheff}
\eea
according to eq.~(\ref{defheff}).
Eqs.~(\ref{delmnu}) and (\ref{delheff}) imply that
$\delta {\cal G}$ should contain a double pole
$(P^0-M_\nu+i\epsilon)^{-2}$ 
in order to generate a nonzero shift of the mass $\delta M_\nu \neq 0$
\cite{ffh}.
$\delta {\cal G}$ contains, however,
only a single pole $(P^0-M_\nu+i\epsilon)^{-1}$, since the state
\bea
\delta O \,
\frac{1}{P^0-H+i\epsilon} \, i Q_B \,
\ket{\vec{p},-\vec{p},\lambda,\bar{\lambda}} 
\label{state}
\eea
in eq.~(\ref{deltag})
does not include the boundstate pole.
This follows from the
physical state condition eq.~(\ref{physstatecond}).
Also one can see explicitly by power countings of diagrams that 
at any order of $1/c$ expansion the above state
does not contain this boundstate pole;
a proof is given in Appendix~\ref{appb}.
Thus, the boundstate mass is gauge independent 
in spite of the fact that
the effective Hamiltonian is gauge dependent.

In addition, the proof in Appendix~\ref{appb} 
also shows that there is no unphysical state which
contributes a pole to the Green function ${\cal G}$ that is
degenerate with or close to one of
the poles of the physical boundstates of our interest.
Stating more explicitly, 
there is no unphysical boundstate with a binding energy
$\sim \alpha_S^2 m$.

Now let us define a quantum mechanical operator
$Q(P^0)$ (which operates only on the subspace of two-body states)
by
\bea
&&
\bra{\vec{p},-\vec{p},\lambda,\bar{\lambda}}
Q (P^0) 
\ket{\vec{q},-\vec{q},\lambda',\bar{\lambda}'}
\nonumber \\ &&
= \int \frac{d^3 \vec{q}\, '}{(2\pi)^3} \sum_{\lambda'',\bar{\lambda}''} \,
\bra{\vec{p},-\vec{p},\lambda,\bar{\lambda}} 
Q_B \, \frac{1}{P^0-H+i\epsilon}
\,  \delta O  \,
\frac{1}{P^0-H+i\epsilon}
\ket{\vec{q}\, ',-\vec{q}\, ',\lambda'',\bar{\lambda}''} 
\nonumber \\ &&
~~~~~~~~~~~~~
~~~~~~~~~~~~~
\times
{\cal G}^{-1} 
(\vec{q}\, ',\vec{q};\lambda'',\bar{\lambda}'',\lambda',\bar{\lambda}';P^0).
\label{defcharge}
\eea
Then $Q(P^0)$ does not include the boundstate poles 
$(P^0-M_\nu+i\epsilon)^{-1}$.
$Q$ can be interpreted as the generator of the transformation
of gauge-fixing condition as seen from the relations
\bea
\delta {\cal G} = - i Q \, {\cal G} + i {\cal G} \, Q^\dagger ,
\label{defq}
\eea
and
\bea
\delta H_{\rm eff} = 
\left[ H_{\rm eff}(P^0) - P^0 \right] \, i Q (P^0)
- i Q^\dagger (P^0) \, \left[ H_{\rm eff}(P^0) - P^0 \right] .
\label{concise}
\eea
The last equation concisely represents the transformation of the effective
Hamiltonian in a form which
clearly shows the spectral invariance; cf.\ eq.~(\ref{delmnu}).
One may easily see that the charge $Q$ has following properties:
in general $Q$ is not Hermite, 
thus the transformation
is non-unitary;
$Q$ vanishes at leading order of the $1/c$ expansion;
beyond leading order, even at some specific order of $1/c$
the charge $Q$ contains
all orders of $\alpha_S$ due to the form of eq.~(\ref{defcharge}).
We will confirm these properties by explicit
calculations in Section~\ref{s7}.

Another method to verify gauge independence of the boundstate spectrum
is as follows.
The on-shell $q\bar{q}$ scattering amplitude can be calculated
using the reduction formula of time-ordered perturbation theory,
\bea
{\cal M}_{q\bar{q}\to q\bar{q}} = 
\lim_{\omega_{\vec{p}},\omega_{\vec{q}} \to P^0/2}
( P^0 - 2 \omega_{\vec{p}} )
( P^0 - 2 \omega_{\vec{q}} )
\,
{\cal G}(\vec{p},\vec{q};\lambda,\bar{\lambda},\lambda',\bar{\lambda}';P^0) 
.
\label{onshell}
\eea
(See Appendix~\ref{app2}.)
If this amplitude is analytically continued to an unphysical
region, it exhibits a pole at the position of the 
boundstate, $P^0= 2\omega_{\vec{p}} = 2\omega_{\vec{q}} \to M_\nu$.
If we expand the amplitude as a Laurent series at the
pole
\bea
{\cal M}_{q\bar{q}\to q\bar{q}} = 
\frac{R_\nu}{P^0-M_\nu+i\epsilon} + (\mbox{regular as }P^0 \to M_\nu) ,
\eea
and calculate the mass $M_\nu$ in a perturbative series in $1/c$,
$M_\nu$ should be gauge independent at each order of $1/c$, since 
${\cal M}_{q\bar{q}\to q\bar{q}}$ 
is gauge independent at any order of perturbation series in $\alpha_S$.

Next we turn to the boundstate wave function, which is defined from 
a Laurent expansion of the Green function at $P^0 = M_\nu$
as\footnote{
At leading order of $1/c$ expansion, $\nu = (n,s,\bar{s})$ and
\bea
\varphi_\nu^{\rm (LO)} (\vec{p},\lambda,\bar{\lambda}) = 
\phi_n (\vec{p}) \, \xi_s(\lambda) \, \xi_{\bar{s}}(\bar{\lambda}) ,
~~~~~ ~~~~~
M_\nu^{\rm (LO)}  = 2 m - \frac{(C_F\alpha_S)^2 m}{4n^2} ,
\nonumber
\eea
where $\xi_s(\lambda) = \braket{\lambda}{s}$
is a two-component spin wave function.
Expressions of $\varphi_\nu$ and $M_\nu$ up to ${\cal O}(1/c^2)$ 
for the boundstates can be
found in \cite{py,analytic}.
}
\bea
{\cal G}(\vec{p},\vec{q};\lambda,\bar{\lambda},\lambda',\bar{\lambda}';P^0)
= \frac{
\varphi_\nu (\vec{p},\lambda,\bar{\lambda}) \,
\varphi_\nu^* (\vec{q},\lambda',\bar{\lambda}')
}{
P^0 - M_\nu + i\epsilon
} 
+ (\mbox{regular as }P^0 \to M_\nu) ,
\eea
or equivalently, 
\bea
\varphi_\nu (\vec{p},\lambda,\bar{\lambda}) = 
\braket{\vec{p},-\vec{p},\lambda,\bar{\lambda}}{\nu;{\rm eff}} 
\eea
with a normalization condition
\bea
\bra{\nu;{\rm eff}} \, 
1 - H_{\rm eff}'(M_\nu) \, \ket{\nu;{\rm eff}} = 1 .
\label{normcond}
\eea
Alternatively, from the original definition of ${\cal G}$, 
eq.~(\ref{defcalg}), one may express
\bea
\varphi_\nu (\vec{p},\lambda,\bar{\lambda}) = 
\braket{\vec{p},-\vec{p},\lambda,\bar{\lambda}}{\nu;\vec{P}=0} ,
\label{projwavefn}
\eea
where $\ket{\nu;\vec{P}=0}$ is the eigenstate of the full QCD
Hamiltonian $H$; see Section~\ref{s2}.
Then from eqs.~(\ref{deltag}) and (\ref{defq}) 
the variation of the wave function
is given by
\bea
\delta \varphi_\nu (\vec{p},\lambda,\bar{\lambda}) &=& 
- \bra{\vec{p},-\vec{p},\lambda,\bar{\lambda}}
i Q_B \, \frac{1}{M_\nu-H+i\epsilon} \, \delta O
\ket{\nu;\vec{P}=0} 
\label{varwavefn}
\eea
and
\bea
\delta \varphi_\nu (\vec{p},\lambda,\bar{\lambda}) &=& 
-i \, [ Q(M_\nu) \cdot \varphi_\nu ] (\vec{p},\lambda,\bar{\lambda}) 
\nonumber \\
&\equiv& 
-i \, \int \frac{d^3 \vec{q}}{(2\pi)^3} \sum_{\lambda',\bar{\lambda}'} \,
\bra{\vec{p},-\vec{p},\lambda,\bar{\lambda}}
Q (M_\nu) 
\ket{\vec{q},-\vec{q},\lambda',\bar{\lambda}'}
\, \varphi_\nu (\vec{q},\lambda',\bar{\lambda}')
\label{varwavefn2}
\eea
when the gauge-fixing condition is varied.
The last equation shows once again that $Q$ can be interpreted
as the transformation charge.

Looking at eq.~(\ref{varwavefn}) one might think that it is possible
to mix different 
gauges in calculations of decay amplitudes of the boundstate.
Namely, one might take the wave function 
$\varphi_\nu (\vec{p},\lambda,\bar{\lambda})$
calculated in one gauge (e.g.\ Coulomb gauge) 
as the initial state wave function and calculate the rest of the
decay amplitude in another gauge (e.g.\ Feynman gauge).
Generally final states satisfy the physical state
condition $Q_B \ket{f} = 0$, so the above equation may suggest
that such a calculation gives the correct result
(the result of a consistent calculation in one specific gauge).
This expectation, however, is wrong since the two-body states
$\ket{\vec{p},-\vec{p},\lambda,\bar{\lambda}}$ do not span the
complete Fock space.
This fact will be verified 
explicitly in the second example in Section~\ref{s7}.

\section{Problems with the On-shell Matching Procedure}
\label{s6}
\clfn

In the definition of the effective Hamiltonian in terms of the full QCD 
Hamiltonian [eqs.~(\ref{defcalg}) and (\ref{defheff})] we kept the
energies of the initial and final $q\bar{q}$ states different from
$P^0$.
(It corresponds to off-shell initial and final states in the
language of a Lorentz covariant formulation.)
Accordingly the form of $H_{\rm eff}$ depends
on our choice of gauge.
In some literatures, however, the on-shell scattering amplitude 
eq.~(\ref{onshell}) is used
instead of the off-shell
Green function ${\cal G}$ in order to determine a similar 
effective Hamiltonian.
This leaves, in general, more freedom to the form of the
effective Hamiltonian than what is due to gauge dependences.
The difference is irrelevant when the Hamiltonian is 
applied to describe an on-shell $q\bar{q}$ system, whereas
the quark and antiquark inside a
boundstate are generally off-shell.
In this section we examine how the spectrum and the wave functions
of the boundstates are affected when we employ the on-shell matching
procedure to determine the effective Hamiltonian.

First we consider a variation of the boundstate mass as we vary
the effective Hamiltonian under the constraint that it gives the same
on-shell scattering amplitude.
As we have seen in eq.~(\ref{delheff}), $\delta {\cal G}$ should include
a double pole $(P^0-M_\nu+i\epsilon)^{-2}$ in order to generate a
nonzero mass shift.
We may try a simplest example:
\bea
\delta {\cal G}
(\vec{p},\vec{q};\lambda,\bar{\lambda},\lambda',\bar{\lambda}';P^0)
= 
\frac{\Delta M_\nu}{(P^0-M_\nu+i\epsilon)^2} 
\, (2\pi)^3\delta^{(3)}(\vec{p}-\vec{q}) \, 
\delta_{\lambda \lambda'} \delta_{\bar{\lambda} \bar{\lambda}'}
,
\\
{\rm i.e.}
~~~~~ ~~~~~
\delta H_{\rm eff}(P^0) 
=
\frac{\Delta M_\nu}{(P^0-M_\nu+i\epsilon)^2} \,
\left[ P^0 - H_{\rm eff}(P^0) \right]^2 .
\label{exheff}
\eea
Evidently it does not modify the on-shell amplitude (\ref{onshell}), 
while it
does generate a mass shift 
$M_\nu \to M_\nu + \Delta M_\nu$.
In the calculation of the boundstate mass in a 
perturbative expansion in $1/c$, if we add the above
$\delta H_{\rm eff}$ 
to the effective Hamiltonian retaining terms
up to some chosen order in $1/c$, the mass is shifted up to the
corresponding order.
In fact one may find a variety of examples which can affect the
boundstate mass while keeping the on-shell amplitude unchanged.
Nevertheless we consider that it will 
not create a serious problem in practice, since
we do not see any good reason why $\delta H_{\rm eff}$ which has explicit
pole structure(s) should mix in the determination of $H_{\rm eff}$.

Next we consider the boundstate wave functions.
Generally the wave function $\varphi_\nu$ changes
when $\delta {\cal G}$ includes a single pole 
$(P^0-M_\nu+i\epsilon)^{-1}$.
For example, if we take
\bea
\delta {\cal G} &=& X \, {\cal G} + {\cal G} \, X ,
\\ 
{\rm i.e.} ~~~~~ ~~~~~
\delta H_{\rm eff} &=& 
[ P^0 - H_{\rm eff}(P^0) ] \, X +
X \, [ P^0 - H_{\rm eff}(P^0) ] ,
\eea
the on-shell amplitude is not affected, 
where $X$ is non-diagonal in momentum space and does not include
the free particle poles 
$(P^0-2\omega_{\vec{p}})^{-1}$, 
$(P^0-2\omega_{\vec{q}})^{-1}$.
On the other hand, the wave function varies as
\bea
\delta \varphi_\nu = X \cdot \varphi_\nu .
\eea
In this case the variation is serious, since different calculations of
a decay amplitude of a boundstate 
do not lead to a unique result
if one uses different $\varphi_\nu$'s connected by the above
transformation as the intial state
wave functions.

One may think that the ambiguity related to the on-shell 
matching procedure to determine $H_{\rm eff}$ can be eliminated
by directly matching all the relevant
on-shell amplitudes to the perturbative expansion of the same amplitudes
in $\alpha_S$.
This works at lower orders of $1/c$ expansions (in Coulomb gauge),
but from the order
$1/c^3$ there appear contributions from the ``ultra-soft gluons'' which 
include all orders of $\alpha_S$ \cite{kp}
such that one should really consider
the off-shell matching procedure seriously.

We conclude, therefore, that the determination of the effective
Hamiltonian $H_{\rm eff}$ from the off-shell Green function
${\cal G}$ is favorable, and that the on-shell matching procedure
can in general 
lead to incorrect calculations of the boundstate masses and the
physical amplitudes involving boundstates.

\section{Examples}
\label{s7}
\clfn

In this section we apply our formalism to two examples,
where we study an infinitesimal gauge transformation from the
Coulomb gauge.
First example is a calculation of the transformation charge $Q$;
in the second example we study gauge dependences of 
diagrams for a decay amplitude of a boundstate.

Let us consider a class of gauge-fixing functions which
interpolates the Coulomb gauge and the Feynman gauge.
The gauge-fixing function is chosen as
\bea
F = -2i \left( \frac{1}{2} B + \partial_\mu A^\mu 
+ \frac{1}{\xi m^2}\, \square \,  \vec{\nabla} \cdot \vec{A} \right) ,
\label{ourgauge}
\eea
from which one obtains
\bea
{\cal L}_{\rm GF+FP} = - {\rm tr}
\left[ \biggl( \partial_\mu A^\mu 
+ \frac{1}{\xi m^2}\, 
{\square} \, \vec{\nabla} \cdot \vec{A} \biggr)^2 \right]
+ 2i \, {\rm tr} \left[
\bar{c} \biggl( \partial_\mu D^\mu 
+ \frac{1}{\xi m^2}\, \square \, \vec{\nabla} \cdot \vec{D} \biggr) c \right]
\eea
after integrating out the auxiliary field $B$.
Here, $\xi>0$ is the gauge parameter:
$\xi \to 0$ and $\xi \to \infty$ correspond to the Coulomb gauge and 
the Feynman gauge, respectively.
The gluon propagator $i D_{\mu \nu}(k)$ is given by
\bea
i D_{00} &=& \frac{-i}{k^2+i\epsilon} \biggl( 1 - \frac{1}{a^2} \biggr)
+ \frac{i}{ | \vec{k} |^2 a^2 } ,
\\
i D_{i0} &=& \frac{i}{k^2+i\epsilon} \, \frac{k^i k^0}{ | \vec{k} |^2 a^2 }
\, \frac{\xi m^2}{ | \vec{k} |^2 } ,
\\
i D_{ij} &=& \frac{i}{k^2+i\epsilon} \biggl( \delta^{ij} -
\frac{k^ik^j}{| \vec{k} |^2 a^2 } \, [ 1 + 2 \xi m^2/| \vec{k} |^2 ]
\biggr) ,
\eea
where $a = 1 + \xi m^2/| \vec{k} |^2$.
Our formal arguments in the previous sections
do not apply directly to this gauge-fixing
condition since $\delta F$ includes $\partial_0 \vec{A}$.
Nevertheless we may obtain necessary rules 
for time-ordered perturbation theory
easily via relations similar to 
eqs.~(\ref{quarkpp})-(\ref{tgluonpp}).\footnote{
A more natural choice of gauge-fixing function that interpolates the
Coulomb gauge and the Feynman gauge would be
$$
F = -2i \left( \frac{1}{2} B + \partial_\mu A^\mu 
+ \frac{1}{\xi} \vec{\nabla} \cdot \vec{A} \right) .
$$
In this case, canonical quantization can be performed 
straightforwardly following the
standard procedure \cite{canonical} and all of our formal arguments apply 
directly.
On the other hand, practical calculations are tediously complicated 
in this gauge due to the
existence of a double pole $(k^2+i\epsilon)^{-2}$ in the gluon propagator.
For simplicity of practical calculations, we present the examples
according to the gauge-fixing condition eq.~(\ref{ourgauge}).
Another class of gauge-fixing conditions which interpolates these two
gauges was introduced, for QED, in \cite{heckathorn}, 
which corresponds to a class of nonlocal gauge-fixing functions.
}
For an infinitesimal change of the parameter $\xi \to \xi + \delta \xi$,
\bea
\delta O = \int d^3 \vec{x} \, \,
\frac{2i\delta \xi }{\xi^2 m^2} \,
{\rm tr} [ \bar{c} \, \square \vec{\nabla} \cdot \vec{A} ] .
\eea

\subsection{The Charge $Q$ at ${\cal O}(1/c)$}

First we calculate the transformation charge $Q$ which generates
an infinitesimal gauge transformation from the
Coulomb gauge ($\xi = 0$) at ${\cal O}(1/c)$.
For perturbative calculations it is convenient to rewrite
eq.~(\ref{defcharge}) as
\bea
Q (P^0) = P \, Q_B \, \frac{1}{P^0-H+i\epsilon} \, 
\delta O \,
\frac{1}{ 1 - \overline{P} \, \frac{1}{P^0-H_0+i\epsilon} \, V}
\, P ,
\label{ptcharge}
\eea
where $H=H_0 + V$.
$P$ denotes the projection operator to the subspace spanned by the
two-body states $\ket{\vec{p},-\vec{p},\lambda,\bar{\lambda}}$, and
$\overline{P} = 1 - P$.
Time-ordered diagrams are obtained by
expanding the above equation in terms of $V$ and inserting the completeness
relation in terms of the eigenstates of $H_0$.
We may discard diagrams without cross talks between
$q\bar{q}$ and ghost sectors, i.e.\ those diagrams which contain
vacuum bubbles.\footnote{
This corresponds to renormalizing the perturbative vacuum
$\ket{0}_{\rm free}$
appropriately in each gauge.
}
The on-shell renormalization scheme is assumed for any value of
$\xi$, so we may neglect quark self-energy diagrams at ${\cal O}(1/c)$.
The BRST charge reads
\bea
Q_B = \int d^3\vec{x} \, \, g \, \psi^\dagger(x) c(x) \psi (x) + \cdots ,
\eea
where only the term which contributes up to ${\cal O}(1/c)$ is shown.

\begin{figure}[tbp]
  \begin{center}
    \includegraphics[width=11cm]{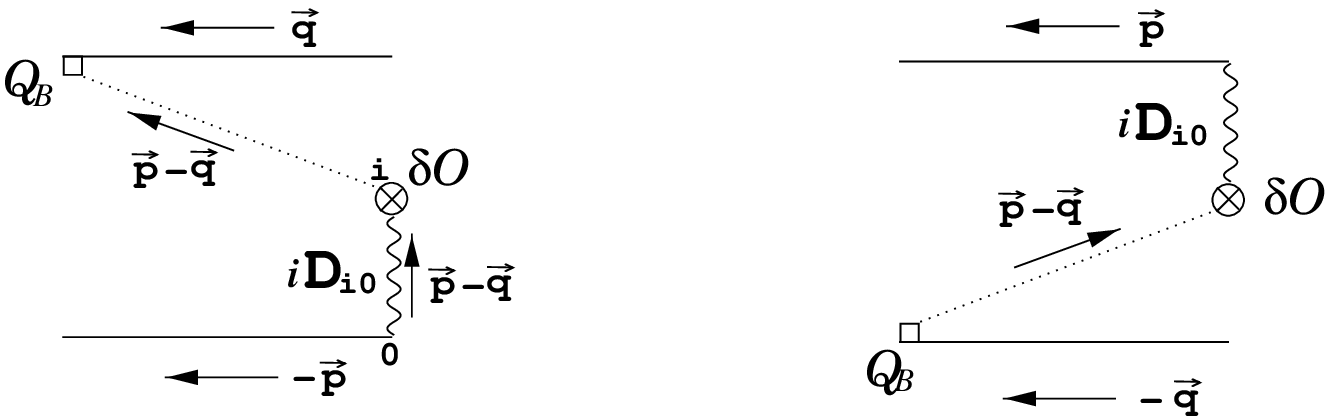}\\
    \vspace{8mm}
    \caption{\small \label{d1}
The tree diagrams which contribute to the charge $Q$ for an
infinitesimal transformation from the Coulomb gauge.
The dotted line represents the ghost.
The wavy line represents 
the gluon propagator $i D_{i0}$; it is reduced effectively to
an instantaneous propagator since the pole $(k^2+i\epsilon)^{-1}$
is cancelled by $k^2$ included in 
$\delta O$.
    }
  \end{center}
\end{figure}

\begin{figure}[tbp]
  \begin{center}
    \includegraphics[width=15cm]{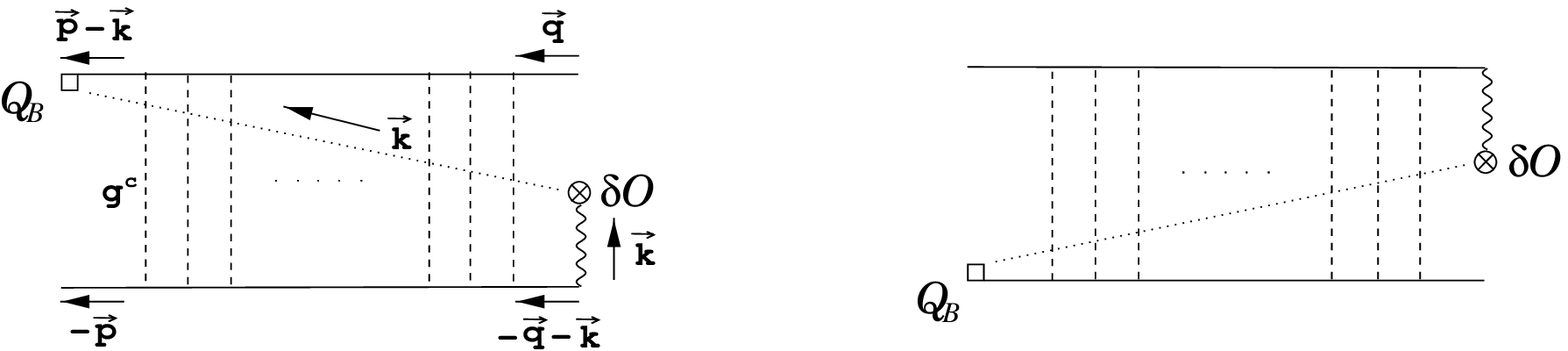}\\
    \vspace{8mm}
    \caption{\small \label{d2}
The diagrams which contribute to the charge $Q$ at ${\cal O}(1/c)$
for an infinitesimal transformation from the Coulomb gauge.
The dashed line represents the Coulomb gluon of the Coulomb gauge.
Other notations are same as in Fig.~\ref{d1}.
    }
  \end{center}
\end{figure}

Simplest diagrams generated by eq.~(\ref{ptcharge}) are the tree diagrams
shown in Fig.~\ref{d1}.
The two diagrams give equal contributions, 
the sum of which is given by
\bea
\bra{\vec{p},-\vec{p},\lambda,\bar{\lambda}}
Q (P^0) 
\ket{\vec{q},-\vec{q},\lambda',\bar{\lambda}'}
\biggr|_{{\cal O}(\alpha_S)}
= i \, C_F \, 4\pi\alpha_S \, \frac{\delta \xi \, m^2}{|\vec{p}-\vec{q}|^4} \,
\frac{1}{P^0-|\vec{p}-\vec{q}|-\omega_{\vec{p}}-\omega_{\vec{q}}+i\epsilon} .
\eea
Examining variations of the boundstate wave functions generated
by this charge [eq.~(\ref{varwavefn2})], we see that two regions of the
gluon-ghost momentum, 
\bea
\begin{array}{lcl}
\mbox{soft}&:& ~~~ |\vec{p}-\vec{q}| \sim {\cal O}(\beta) ,
\\
\mbox{ultra-soft}&:& ~~~ |\vec{p}-\vec{q}| \sim {\cal O}(\beta^2) ,
\end{array}
\eea
are relevant at ${\cal O}(1/c)$.\footnote{
In power counting we consider
$\delta \xi \, m^2/|\vec{p}-\vec{q}|^2 \sim {\cal O}(1)$.
}
Existence of the ultra-soft region indicates that the diagrams
with multiple Coulomb-gluon exchange in ladder 
contribute also at ${\cal O}(1/c)$.
Indeed one may check that all the diagrams shown in
Fig.~\ref{d2} contribute to $\delta \varphi_\nu$
at this order;
the contributions come from the ultra-soft region
of the gluon-ghost momentum, 
$|\vec{k}| \sim {\cal O}(\beta^2)$. 
This is also consistent with
the result of Love \cite{love}.
Hence, we find
\bea
&&
\bra{\vec{p},-\vec{p},\lambda,\bar{\lambda}}
Q (P^0) 
\ket{\vec{q},-\vec{q},\lambda',\bar{\lambda}'}
\biggr|_{{\cal O}(1/c)}
\nonumber \\
&&= i \, C_F \, 4\pi\alpha_S \, \int \frac{d^3\vec{k}}{(2\pi)^3}
\, \frac{\delta \xi \, m^2}{|\vec{k}|^4} \,
{\cal G}^{\rm (LO)}
\biggl( {\textstyle
\vec{p}-\frac{\vec{k}}{2},\vec{q}+\frac{\vec{k}}{2};
\lambda,\bar{\lambda},\lambda',\bar{\lambda}';
P^0-|\vec{k}|-\frac{|\vec{k}|^2}{4m} 
} \biggr) ,
\label{nlocharge}
\eea
where
\bea
{\cal G}^{\rm (LO)}
( \vec{p},\vec{q};\lambda,\bar{\lambda},\lambda',\bar{\lambda}';P^0 ) 
= \bra{\vec{p},-\vec{p},\lambda,\bar{\lambda}}
\frac{1}
{P^0-H^{\rm (LO)}_{\rm eff}+i\epsilon} 
\ket{\vec{q},-\vec{q},\lambda',\bar{\lambda}'} 
\eea
includes summation of the Coulomb ladders to all orders of $\alpha_S$.
The charge $Q(P^0)$ turns out to be anti-hermite at ${\cal O}(1/c)$.
We note that the above charge does not include any boundstate pole
because of the integration over $\vec{k}$.

Alternatively it is possible to calculate the charge
$Q|_{{\cal O}(1/c)}$ by first evaluating 
$\delta H_{\rm eff}$ and
then extracting $Q$ via the relation eq.~(\ref{concise}).
This procedure becomes cumbersome at higher orders of
$\alpha_S$ because the number of gauge cancellations among 
diagrams increases.
These gauge cancellations are automatically incorporated
in the direct calculation of $Q$ above by the BRST
invariance of the full QCD Hamiltonian,
$[ Q_B, H ] = 0$.

\subsection{A Decay Amplitude of a $q\bar{q}$ Boundstate at 
${\cal O}(1/c)$}

\begin{figure}[tbp]
  \begin{center}
    \includegraphics[width=13cm]{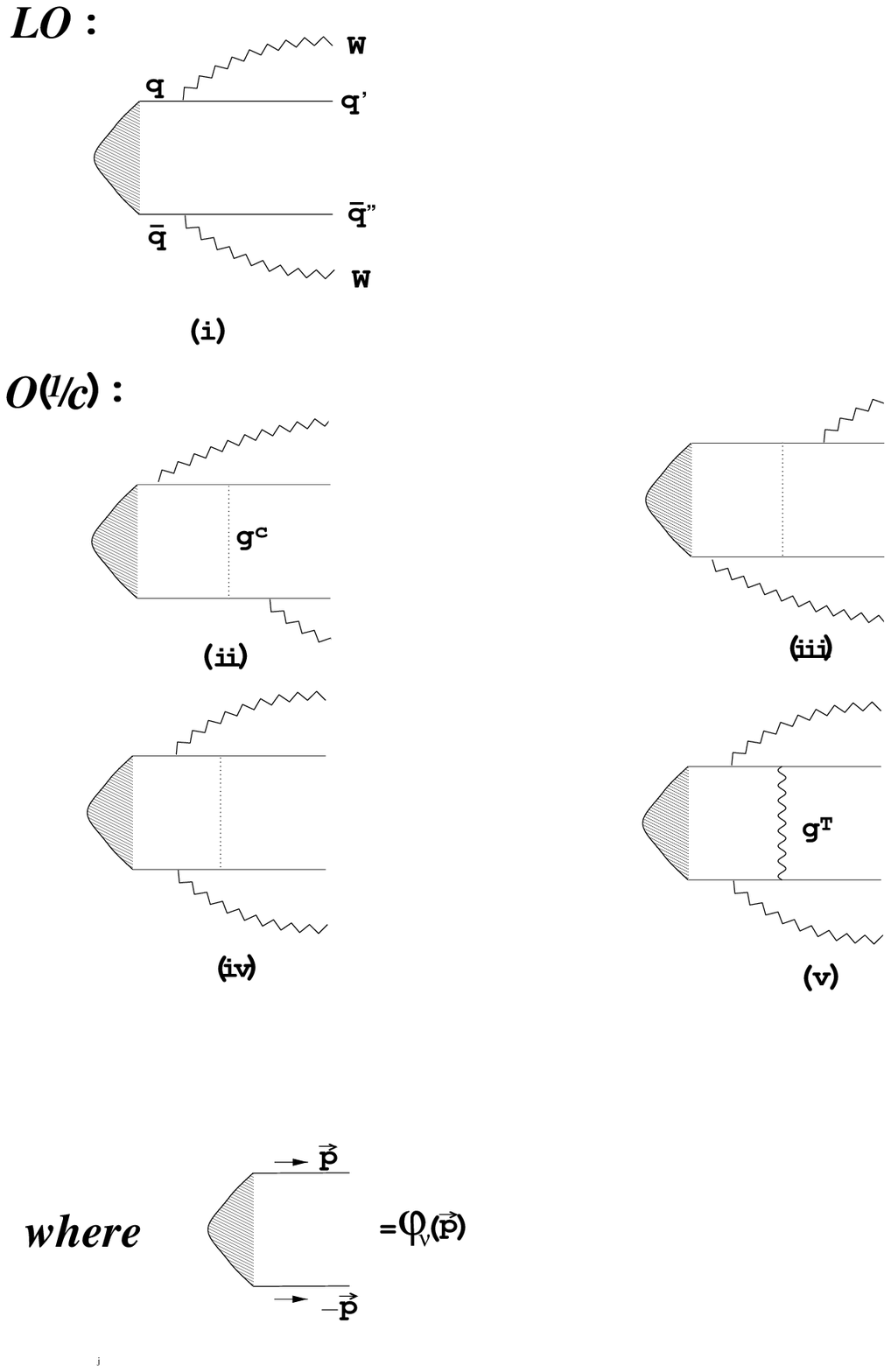}\\
    \vspace{8mm}
    \caption{\small \label{d3}
The diagrams which contribute to the amplitude for a nonrelativistic
boundstate decaying into $q' \bar{q}'' W^+W^-$ up to ${\cal O}(1/c)$.
We suppressed diagrams for ${\cal O}(\alpha_S)$ corrections
to the $qq'W$ and $\bar{q}\bar{q}''W$ vertices.
    }
  \end{center}
\end{figure}

\begin{figure}[tbp]
  \begin{center}
    \includegraphics[width=11cm]{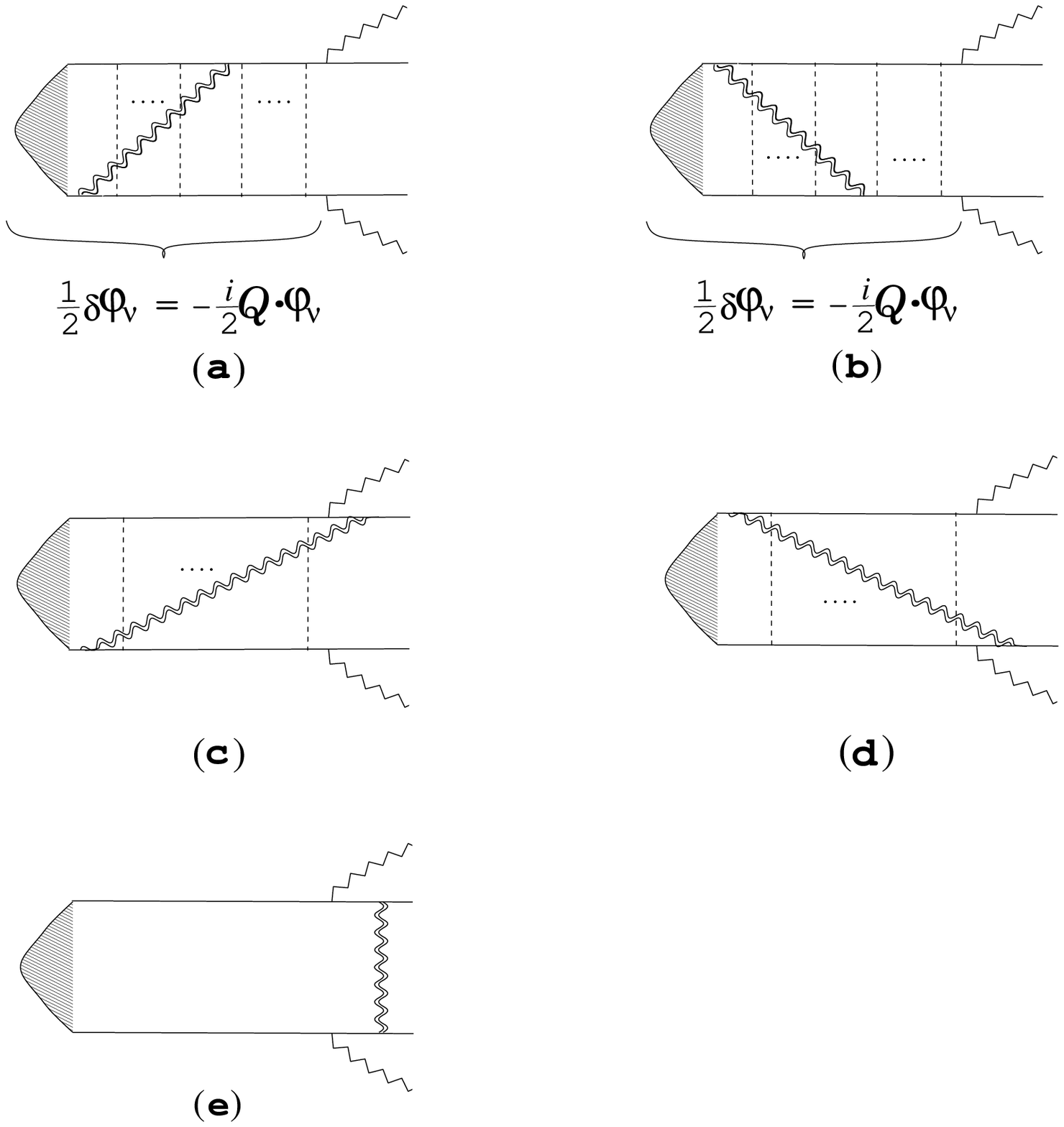}\\
    \vspace{8mm}
    \caption{\small \label{d4}
The diagrams which are generated by the infinitesimal variation
of the gauge-fixing condition from the Coulomb gauge.
The double-wavy line represents the variation of the
gluon propagator $i \, \delta D_{\mu \nu}$.
    }
  \end{center}
\end{figure}

Next we analyze infinitesimal gauge transformations of the
diagrams for the decay process of the boundstate
where $q$ and $\bar{q}$ decay into lighter quarks via
electroweak interaction.
We analyze the infinitesimal transformation from the Coulomb
gauge up to ${\cal O}(1/c)$ as in the above example and
see how the variation of the initial-state wave function
eq.~(\ref{varwavefn2}) gets cancelled in the total amplitude.
The diagrams which contribute to this
process up to ${\cal O}(1/c)$ in Coulomb gauge
are shown in Fig.~\ref{d3} \cite{fsi}.\footnote{
It is understood that the boundstate wave function $\varphi_\nu$
includes ${\cal O}(1/c)$ corrections.
For simplicity, we neglect ${\cal O}(\alpha_S)$ corrections
to the $qq'W$ and $\bar{q}\bar{q}''W$ vertices, which constitute
gauge independent subsets by themselves and do not mix with the
gauge transformation of $\varphi_\nu$.
}
When we vary the gauge-fixing function, additional diagrams
which contribute to the ${\cal O}(1/c)$ decay amplitude
are shown in Fig.~\ref{d4}.
Here, the double-wavy lines represent $i \, \delta D_{\mu \nu}$, where 
\bea
i \, \delta D_{00} = - i \left( \frac{1}{k^2+i\epsilon}+\frac{1}{|\vec{k}|^2} 
\right) \frac{2 \delta \xi \, m^2}{|\vec{k}|^2} ,
\label{deld00}
\eea
etc.
Diagrams~(a) and (b) can be regarded as transformations of the
initial-state wave function of the leading-order diagram~(i).
Conversely, the diagrams (c)-(e) cannot be regarded as such,
since they do not contain a $q\bar{q}$ two-body state
as an intermediate state.

\begin{figure}[tbp]
  \begin{center}
    \includegraphics[width=15cm]{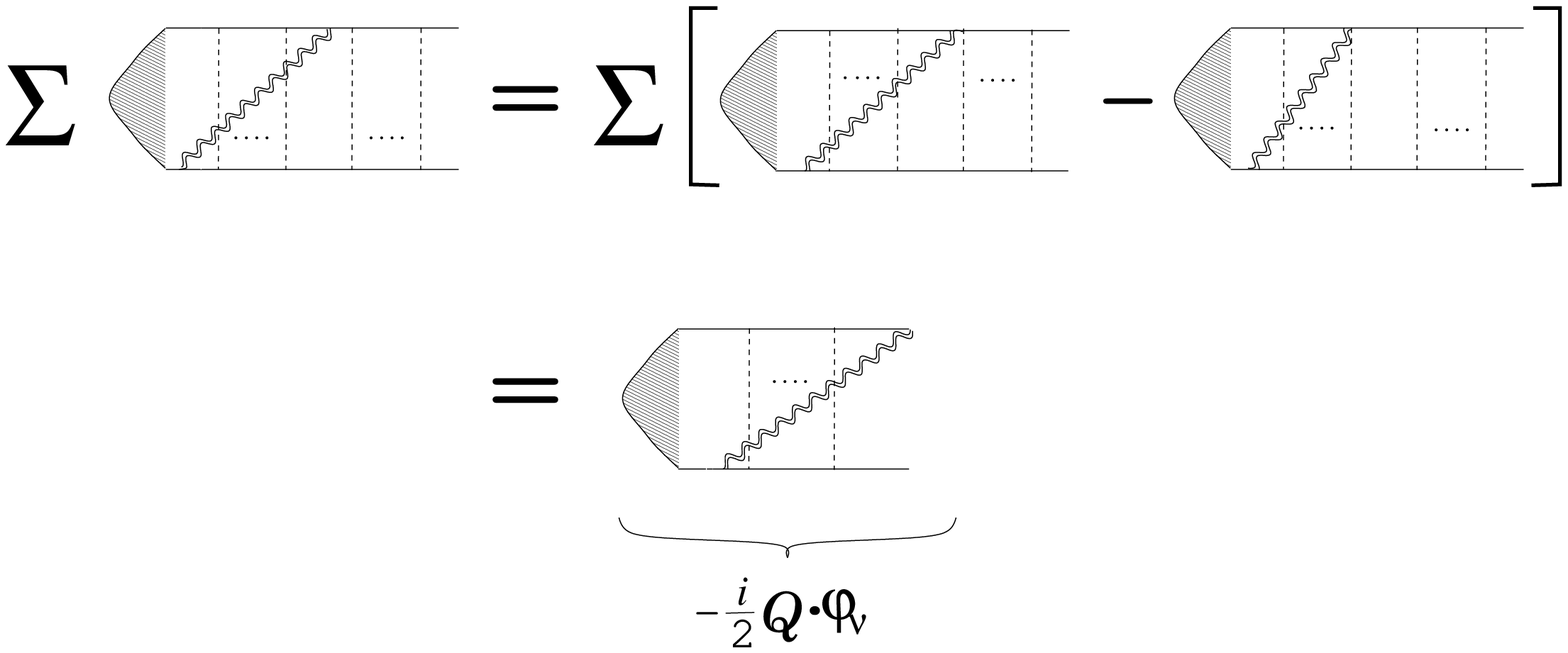}\\
    \vspace{8mm}
    \caption{\small \label{d5}
{\protect
A diagrammatic method to show that the sum of diagrams in 
Fig.~\ref{d4}(a) can be regarded as the leading order diagram
Fig.~\ref{d3}(i) with the initial state boundstate wave function
$\varphi_\nu$
replaced by a half of its infinitesimal transformation 
$-iQ\cdot \varphi_\nu/2$.
    }}
  \end{center}
\end{figure}

\begin{figure}[tbp]
  \begin{center}
    \includegraphics[width=17cm]{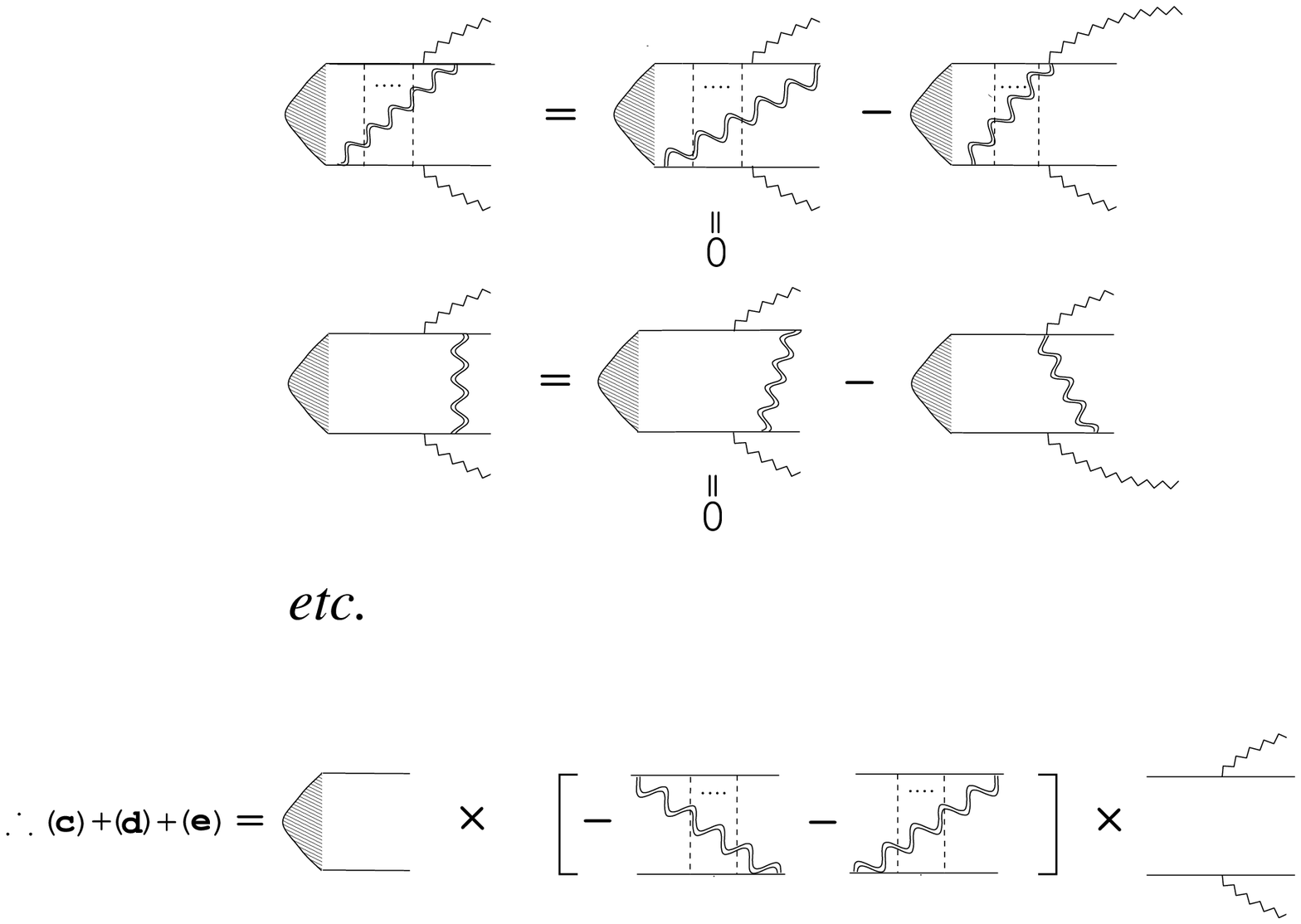}\\
    \vspace{8mm}
    \caption{\small \label{d6}
{\protect
Examples of the diagrammatic method to show that the sum of
diagrams in Fig.~\ref{d4}(c)-(e) exactly cancels the sum of (a) and 
(b).
    }}
  \end{center}
\end{figure}

Using diagrammatic analysis, one may
verify that the sum of all diagrams ~(a)-(e) vanishes so that the
total amplitude is indeed gauge independent.
In fact, from diagrammatic manipulations
as shown in Fig.~\ref{d5} and also from similar manipulations corresponding to
the diagram~(b), one can show that the sum of diagrams in (a) and (b)
can be regarded as the leading order diagram
Fig.~\ref{d3}(i) with the initial state boundstate wave function
$\varphi_\nu$
replaced by its infinitesimal transformation eq.~(\ref{varwavefn2}).
Rearrangement of diagrams may be performed, for instance, using the
relation
\bea
\frac{1}{P^0-2\omega_{\vec{p}}+i\epsilon} \,
\left( 
\frac{1}{|\vec{k}|}\, 
\frac{1}{P^0-|\vec{k}|-\omega_{\vec{p}}-\omega_{\vec{p}+\vec{k}}+i\epsilon}
+\frac{1}{|\vec{k}|^2} 
\right) \frac{2 \delta \xi \, m^2}{|\vec{k}|^2} \,
\frac{1}{P^0-2\omega_{\vec{p}+\vec{k}}+i\epsilon} \,
\nonumber \\
= 
\frac{1}{|\vec{k}|^2
(P^0-|\vec{k}|-\omega_{\vec{p}}-\omega_{\vec{p}+\vec{k}}+i\epsilon)} \,
\frac{\delta \xi \, m^2}{|\vec{k}|^2} 
\left(
\frac{1}{P^0-2\omega_{\vec{p}}+i\epsilon} 
+ \frac{1}{P^0-2\omega_{\vec{p}+\vec{k}}+i\epsilon} 
\right)
\eea
for manipulating the propagator eq.~(\ref{deld00}).
On the other hand, from Fig.~\ref{d6} we see that the sum of the diagrams
in (c)-(e) exactly cancels the sum of (a) and (b).
For details of the diagrammatic analyses, see Refs.~\cite{love,ffh}.

According to the formal arguments in Section~\ref{s5} we know how the
initial-state wave function transforms 
and therefore we know the sum of the other diagrams
(c)-(e) in order to ensure gauge independence of the
total amplitude.
This example demonstrates that the diagrammatic analyses are 
rather cumbersome since infinitely many diagrams contribute
even at the lowest nontrivial order of the $1/c$ expansion.

\section{Conclusion and Discussion}
\label{s8}
\clfn

In this paper we analyzed gauge dependence of an effective
Hamiltonian formalism that describes the nonrelativistic quark-antiquark
boundstates and discussed problems of the on-shell matching procedure 
within this formalism.
The significance of our present work may be put as follows.

We used the BRST symmetry, which is known to be a powerful tool
to study QCD Green functions, to analyze the NRQCD boundstates.
The arguments were supplemented by power countings
of singularities of relevant diagrams to make them
more explicit and detailed.
Gauge dependence of the NRQCD boundstate formalism is more
complex than that of usual (naive) perturbation theory
since we have to deal with an infinite number of diagrams at
each order of the $1/c$ expansion.
(e.g. An infinite number of diagrams contribute to
$H_{\rm eff}$ at ${\cal O}(1/c)$ 
in gauges other than the Coulomb gauge \cite{love}.)
We defined $H_{\rm eff}$ naturally in the context of time-ordered
perturbation theory.
Then we obtained
the transformation charge $Q$ of $H_{\rm eff}$,
from which we could easily see gauge independence of the spectrum
and obtain transformation of the boundstate wave functions.
For an infinitesimal transformation from the Coulomb gauge,
we calculated $Q$ directly up to ${\cal O}(1/c)$.
Also we saw that, without resort to the BRST symmetry, 
cumbersome gauge cancellations among diagrams are necessary
to show gauge independence of a decay amplitude of the boundstate.
At higher orders of $1/c$, diagrammatic analyses such as what we 
presented in the second example or those in 
Refs.~\cite{love,ffh} become quite intricate
so that the arguments based on
the BRST symmetry would become more important.

Furthermore, we showed possibilities for incorrect calculations of
amplitudes involving boundstates
if one uses only the on-shell $q\bar{q}$ scattering
amplitude to determine $H_{\rm eff}$.
These problems do not occur 
if we determine $H_{\rm eff}$ from the off-shell Green
function ${\cal G}$, or,
if we use a local NRQCD Lagrangian 
consistently and determine its coefficients via proper
matching procedure, e.g.\ as in lattice calculations
\cite{lattice1,lattice2}.
The latter procedure has a disadvantage that
one should calculate a number of 
amplitudes to determine all the coefficients.

Presently we still do not have at our disposal a
completely systematic way to
identify all the necessary contributions in
computations of physical quantities of the
NRQED/NRQCD boundstates at a given order of $1/c$ expansion.
We believe that the formalism developed in this paper will provide 
useful cross checks in these computations.
Now we know how a boundstate wave
function or the Green function ${\cal G}$ contained in an amplitude
transforms.
The transformation charge $Q$ is process independent
and depends only on the gauge-fixing condition, and
it can be calculated directly in a perturbative expansion in $1/c$.

A possible application is to use the formalism
to study gauge dependences of the diagrams involved in
the calculation of the top quark momentum distribution in
the $t\bar{t}$ threshold region at ${\cal O}(1/c^2)$.
It is known that at leading order the top momentum distribution
is proportional to the absolute square of
the wave functions of (would-be) toponium boundstates in momentum 
space \cite{sumino}.
As we saw in this paper, wave functions of boundstates are 
gauge dependent beyond leading order.
In the second example of Section~\ref{s7},
we verified that this gauge dependence is cancelled by that of 
the final-state interaction diagrams (ii)-(v) at ${\cal O}(1/c)$.
In other words, a boundstate wave function mixes with the final-state
interaction diagrams by gauge transformation.
This shows that the present calculations of the top momentum 
distribution \cite{momdist}
are gauge dependent, i.e.\ they vary if we transform the gauge infinitesimally
from the Coulomb gauge, since they do not include the final-state interaction
diagrams.
Also the example suggests how
gauge cancellations should take place in the complete amplitude at
${\cal O}(1/c^2)$ which has not been obtained yet.

\section*{\bf Acknowledgements}

We would like to thank 
K.~Hikasa and K.~Sasaki for fruitful discussions.
One of the authors (Y.S.) is grateful to A.~Czarnecki and
T.~Onogi for discussions.
This work was supported by the Japan-German Cooperative Science
Promotion Program.

\begin{appendix}

\section{Use of the Equation of Motion in a Local NRQCD Lagrangian}
\label{appa}
\clfn

In writing down a local NRQCD Lagrangian in terms of the 
nonrelativistic quark ($\psi_q$), nonrelativistic antiquark 
($\psi_{\bar{q}}$), 
gluon ($A_\mu$), ghost ($c$) and antighost ($\bar{c}$)
fields, in principle one writes down all possible local interactions 
consistent with the rotational and BRST symmetries.
In addition, one may simplify the Lagrangian using the equation of
motion, and it is often convenient to
eliminate all terms including $D_0^n~(n \ge 2)$, 
where $D_\mu = \partial_\mu - ig A_\mu (x)$
is the covariant derivative.
After such a simplification, the Lagrangian takes a standard form:
\bea
&&
{\cal L} = \psi_q^\dagger (x) \biggl[
iD_0 + c_2 \frac{\vec{D}^2}{2m} + c_4 \frac{\vec{D}^4}{8m^3} +
c_F \frac{g}{2m} \vec{B}\cdot\vec{\sigma} +
c_D \frac{g}{8m^2} \left( \vec{D}\cdot\vec{E}-\vec{E}\cdot\vec{D}\right)
\nonumber \\ && ~~~~~ ~~~~~
+ c_S \frac{g}{8m^2} i \vec{\sigma}\cdot \left(\vec{D}\times\vec{E}
-\vec{E}\times\vec{D} \right) + \cdots
\biggr] \psi_q (x) + (\psi_q \to \psi_{\bar{q}})
\nonumber \\ && ~~~~~
+ c_{\mbox{\scriptsize {\it 4-Fermi} }} \, \frac{g^2}{m^2} 
  \psi_q^\dagger (x) \psi_{\bar{q}}^\dagger (x) 
\psi_q (x) \psi_{\bar{q}} (x) + \cdots
\nonumber \\ && ~~~~~
- \frac{1}{2} {\rm tr} \left[ G^{\mu\nu} G_{\mu\nu} \right] .
\eea
We suppressed the gauge-fixing and ghost terms.
One should determine the (Wilson) coefficients of local operators
$c_2$, $c_4$, $c_F$, etc.\ by matching various on-shell amplitudes to 
those of full QCD.
Furthermore, in practical applications of the NRQCD formalism, we often
evaluate the correlators involving the current operators
composed of the nonrelativistic quark and/or antiquark fields.
The equation of motion is also used to eliminate $D_0^n$
from the current operators, and
the coefficients of local operators constituting the
current operators are determined by matching the on-shell amplitudes to 
those of full QCD.

In this appendix we prove that we may use the equation of
motion appropriately in order to simplify the form of the Lagrangian.
We also prove that in the evaluation of on-shell amplitudes
involving current operators,
the change of the Lagrangian can be compensated
by local redefinitions of the current operators and that
one can use the equation of motion to rewrite
the current operators.
It is understood that we regularize ultraviolet and infrared divergences 
using the dimensional regularization.

\begin{figure}[tbp]
  \begin{center}
    \includegraphics[width=16cm]{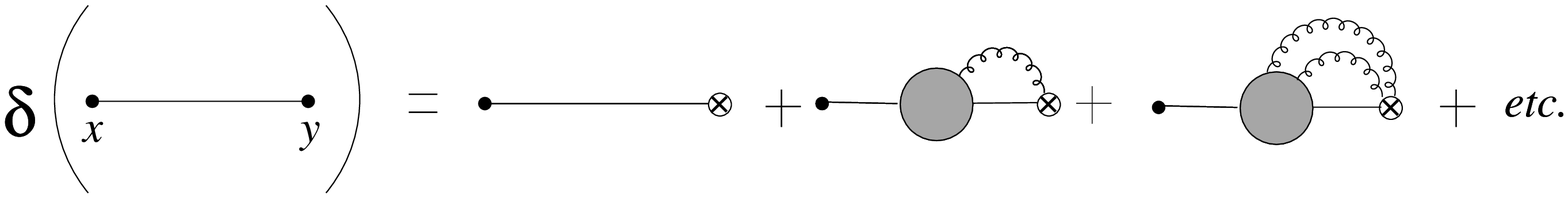}\\
    \vspace{8mm}
    \caption{\small \label{figa1}
{\protect
The diagrammatic representation of eq.~(\ref{wvfnrn}).
$\otimes$ shows the position of the local operator $N$.
The pole position is not changed, while the $Z$ factor changes.
    }}
  \end{center}
\end{figure}

Let us start from a general local Lagrangian 
${\cal L}(\psi_q, \psi_{\bar{q}}, A_\mu, c, \bar{c})$
and add a local operator which vanishes by the equation of motion:
\bea
{\cal L} \to {\cal L} + \psi^\dagger_{q} \{ N , M \} \psi_q .
\label{varlag}
\eea
Here, the equation of motion for $\psi_q$ is denoted by
\bea
\frac{\delta S}{\delta \psi^\dagger_q (x)} = \left( M \psi_q \right) (x) ,
~~~~~ ~~~~~
S = \int d^D x \, {\cal L} ,
\eea
and $N$ denotes a local operator with $N=N^\dagger$, e.g.\ 
$N \, \psi_q = i D_0 \psi_q$, $\vec{D}\, ^2 \psi_q$,
$B \cdot \sigma  \, \psi_q$, etc.
$N$ may include the gluon field but not the quark or antiquark
field.
For simplicity we do not change the antiquark sector 
of the Lagrangian in our argument.
According to eq.~(\ref{varlag}),
the two-point and four-point functions change as
\bea
&&
\delta \bra{0} T \, \psi_q(x) \, \psi_{q}^\dagger (y) \ket{0}
= \bra{0} T \, \psi_q(x) \, \psi_{q}^\dagger (y)
\left[ 
i \int d^D z \, \psi^\dagger_{q}(z) ( \{ N , M \} \psi_q )(z) 
\right]
\ket{0} ,
\label{del2ptfn}
\\
&&
\delta \bra{0} T \, \psi_q(x) \, \psi_{q}^\dagger (y)
\, \psi_{\bar{q}} (x') \, \psi_{\bar{q}}^\dagger (y') \ket{0}
\nonumber \\
&& ~~~~~ ~~~~~ ~~~~~ 
= \bra{0} T \, \psi_q(x) \, \psi_{q}^\dagger (y)
\, \psi_{\bar{q}}(x') \, \psi_{\bar{q}}^\dagger (y')
\left[ 
i \int d^D z \, \psi^\dagger_{q}(z) ( \{ N , M \} \psi_q )(z) 
\right]
\ket{0} .
\eea
In order to rewrite the right-hand-side of eq.~(\ref{del2ptfn}) 
one may use the Schwinger-Dyson equation\footnote{
In the path-integral formulation, this follows readily from
$$
\int {\cal D}\psi_q^\dagger \,
\frac{\delta}{\delta \psi_q^\dagger (z)} \,
\psi_q (x) \, \psi_q^\dagger (y) \, [ (N \psi_q)(z) ]^\dagger \, e^{iS} 
= 0
$$
and a similar term with $\psi \leftrightarrow \psi^\dagger$.
}
\bea
&&
\bra{0} T \, \biggl\{
\frac{1}{2} \, \psi_q(x) \, [ (N \psi_{q})(y) ]^\dagger +
\frac{1}{2} \, ( N \psi_q )(x) \, \psi_{q}^\dagger (y) +
\psi_q(x) \, \psi_{q}^\dagger (y) 
\left[ i \int d^D z \, N(z,z) \right] 
\nonumber \\ && ~~~~~
+ \psi_q(x) \, \psi_{q}^\dagger (y)
\left[ 
i \int d^D z \, \psi^\dagger_{q}(z) ( \{ N , M \} \psi_q )(z) 
\right]
\biggr\}
\ket{0}
=0 .
\eea
The third term of this equation vanishes within the dimensional regularization,
since $N(z,z)$ contains $\delta^D (0)$ and/or 
$[ \partial_z^n \delta^D(z) ]_{z\to 0}$ which give scaleless integrals
(tadpoles).
Hence, we have 
\bea
\delta \bra{0} T \, \psi_q(x) \, \psi_{q}^\dagger (y) \ket{0}
= - \bra{0} T \, \biggl\{
\frac{1}{2} \, \psi_q(x) \, [ (N \psi_{q})(y) ]^\dagger +
\frac{1}{2} \, ( N \psi_q )(x) \, \psi_{q}^\dagger (y) 
\biggr\}
\ket{0}.
\label{wvfnrn}
\eea
This equation shows that the change of the Lagrangian does not affect
the pole mass of the quark propagator, whereas the
$Z$-factor (wave function renormalization constant) varies;
see Fig.~\ref{figa1}.

\begin{figure}[tbp]
  \begin{center}
    \includegraphics[width=16cm]{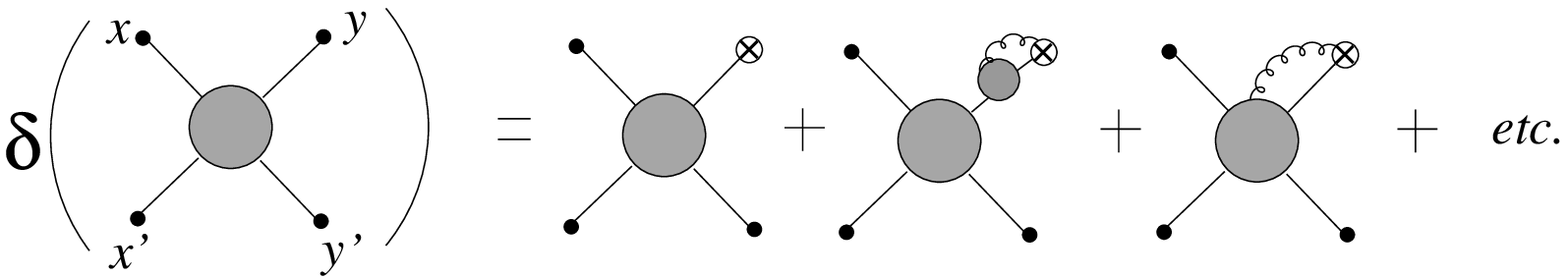}\\
    \vspace{8mm}
    \caption{\small \label{figa2}
{\protect
The diagrammatic representation of eq.~(\ref{del4ptfn2}).
The first two diagrams give rise to a wave function renormalization
common to Fig.~\ref{figa1}.
The third diagram is one-particle irreducible with respect to
the leg with $\otimes$, hence it does not contribute to the
on-shell amplitude.
    }}
  \end{center}
\end{figure}

Following similar steps, one can show that the variation of 
the four-point function is given by
\bea
&&
\delta \bra{0} T \, \psi_q(x) \, \psi_{q}^\dagger (y) 
\, \psi_{\bar{q}} (x') \, \psi_{\bar{q}}^\dagger (y') 
\ket{0}
\nonumber \\ &&
= - \bra{0} T \, \biggl\{
\frac{1}{2} \, \psi_q(x) \, [ (N \psi_{q})(y) ]^\dagger +
\frac{1}{2} \, ( N \psi_q )(x) \, \psi_{q}^\dagger (y) 
\biggr\}
\, \psi_{\bar{q}} (x') \, \psi_{\bar{q}}^\dagger (y') 
\ket{0}.
\label{del4ptfn2}
\eea
Thus, if we redefine the $Z$-factor according to 
eq.~(\ref{wvfnrn}), the on-shell amplitude of the
quark-antiquark scattering remains the same; see Fig.~\ref{figa2}.
Similarly the amplitudes where
multiple gluons are attached to the quark-antiquark scattering 
can be shown to be invariant under the variation of the Lagrangian
eq.~(\ref{varlag}).

Eqs.~(\ref{wvfnrn}) and (\ref{del4ptfn2}) also show 
that, when evaluating correlators involving current operators, 
the change of the
Lagrangian can be compensated by local redefinitions of the
current operators.
By way of example, for a current operator which creates and
annihilates a quark-antiquark pair,
\bea
J^i (x) = \psi_q^\dagger (x)
\left[
c_1^v \sigma^i + c_2^v \sigma^i 
\frac{\stackrel{\leftrightarrow}{D^2}}{12m^2}
+ \cdots
\right] \psi^\dagger_{\bar{q}} (x)  + {\rm h.c.} ,
\eea
the on-shell amplitude calculated from the correlator
$\bra{0} T \, J^i (x) \, 
\psi_q (y) \psi_{\bar{q}}(z) \ket{0}$
remains unchanged if we redefine the current as
\bea
&&
J^i(x) \to J^i(x) + \delta J^i(x) ,
\\
&&
\delta J^i(x) = [ (N \psi_q)(x) ]^\dagger
\left[
c_1^v \sigma^i + c_2^v \sigma^i 
\frac{\stackrel{\leftrightarrow}{D^2}}{12m^2}
+ \cdots
\right] \psi^\dagger_{\bar{q}} (x) + {\rm h.c.} 
\eea

Finally we show that we may use the equation of motion in
order to rewrite the current operators.
One may derive the Schwinger-Dyson equation\footnote{
This follows from
$$
\int {\cal D}\psi \,
\frac{\delta}{\delta \psi_q (z)} \, \Gamma^{i\dagger} (x,z) \,
\psi_{\bar{q}}^\dagger (x) \, \psi_q (y) \, \psi_{\bar{q}}(z) \, e^{iS} 
= 0
$$
and integrating over $z$.
}
\bea
&&
\bra{0} T \, \biggl\{
[ (\Gamma^i M  \psi_{q})(x) ]^\dagger \psi^\dagger_{\bar{q}}(x) 
\psi_q(y) \, \psi_{\bar{q}} (z) 
+ i \Gamma^{i\dagger} (x,y) \, \psi_{\bar{q}}^\dagger(x) \, 
\psi_{\bar{q}} (z)
\biggr\}
\ket{0}
=0 ,
\eea
where $\Gamma^i$ is a local operator and
may include the gluon field but not the quark or
antiquark field, e.g.\ $\Gamma^i(x,y) = \sigma^i \delta^D (x-y)$,
$D^i (x,y)$, etc.
The second term does not contain the quark pole, hence it does not
contribute to the on-shell amplitude.
Thus, adding
$[ (\Gamma^i M  \psi_{q})(x) ]^\dagger \psi_{\bar{q}}^\dagger(x)
+ {\rm h.c.}$
to the current operator $J^i(x)$
does not affect the on-shell amplitude.

\section{Time-ordered Perturbation Theory}
\label{app2}
\clfn

Here, we derive the rules for calculations of
the on-shell quark-antiquark
scattering amplitude in time-ordered (old-fashioned)
perturbation theory.
The $S$-matrix element between the eigenstates of the
free Hamiltonian defined in 
eq.~(\ref{asymstate}) with 
an infinite time separation (asymptotic states)
is given by
\bea
S_{fi} &=& \lim_{T \to \infty}
\, \bra{\vec{p},-\vec{p},\lambda,\bar{\lambda}} 
e^{-iHT}
\ket{\vec{q},-\vec{q},\lambda',\bar{\lambda}'} 
\\ & = & 
\lim_{T \to \infty}
\, \oint \frac{dP^0}{2\pi i} \, e^{-i P^0 T}
\, \bra{\vec{p},-\vec{p},\lambda,\bar{\lambda}} 
\, \frac{1}{P^0-H+i\epsilon} 
\ket{\vec{q},-\vec{q},\lambda',\bar{\lambda}'} .
\eea
In the integrand, we see the Green function
${\cal G}(\vec{p},\vec{q};\lambda,\bar{\lambda},\lambda',\bar{\lambda}';P^0)$
introduced in eq.~(\ref{defcalg}).
We expand the right-hand-side in $V$, where
$H = H_0 + V$,
\bea
\frac{1}{P^0-H+i\epsilon} =
\frac{1}{P^0-H_0+i\epsilon} \sum_{n=0}^\infty
\left( V \, \frac{1}{P^0-H_0+i\epsilon} \right)^n ,
\eea
and insert the completeness relations in terms of the eigenstates of
$H_0$. 
One readily sees that,
at each order of the perturbative expansion, the free propagator poles
$( P^0 - 2 \omega_{\vec{p}} + i\epsilon )^{-1}$ and
$( P^0 - 2 \omega_{\vec{q}} + i\epsilon )^{-1}$
are attached at the both ends.
Therefore, if we write
\bea
\bra{\vec{p},-\vec{p},\lambda,\bar{\lambda}} 
\, \frac{1}{P^0-H+i\epsilon} 
\ket{\vec{q},-\vec{q},\lambda',\bar{\lambda}'} 
= \frac{ {\cal M}_{q\bar{q}\to q\bar{q}}(P^0) }
{ ( P^0 - 2 \omega_{\vec{p}} + i\epsilon )
( P^0 - 2 \omega_{\vec{q}} + i\epsilon ) } 
\eea
and set $2 \omega_{\vec{p}} = 2 \omega_{\vec{q}} = \sqrt{s}$,
we find
\bea
S_{fi} &=& \lim_{T \to \infty}
\, \oint \frac{dP^0}{2\pi i} \, e^{-i P^0 T}
\, \left( \frac{1}{P^0-\sqrt{s}+i\epsilon} \right)^2
{\cal M}_{q\bar{q}\to q\bar{q}}(P^0) .
\eea
Suppose ${\cal M}_{q\bar{q}\to q\bar{q}}(P^0)$ is regular
inside the integration contour.
Then
\bea
S_{fi} &=& \lim_{T \to \infty}
\, \frac{\partial}{\partial P^0} \, 
\{ e^{-i P^0 T} \, {\cal M}_{q\bar{q}\to q\bar{q}}(P^0) \}
\biggr|_{P^0 \to \sqrt{s}}
\\
&=&  \lim_{T \to \infty}
\, e^{-i \sqrt{s} T} 
\{ 
{\cal M}_{q\bar{q}\to q\bar{q}}'(\sqrt{s})
- i T \, {\cal M}_{q\bar{q}\to q\bar{q}}(\sqrt{s})
\} .
\label{95}
\eea
The second term in the last line
represents the dominant term as $T \to \infty$.
Taking into account the irregularity of
${\cal M}_{q\bar{q}\to q\bar{q}}(P^0)$,
we obtain additional terms which are subleading in
comparison to the second term of eq.~(\ref{95}) as
$T \to \infty$.
Thus, we obtain the reduction formula eq.~(\ref{onshell}) as well as
the rules for calculations of the scattering amplitude in 
time-ordered perturbation theory, as explained
in Sec.~\ref{s5}.\footnote{
The phase factor $ e^{-i \sqrt{s} T} $ always appears
in a perturbative evaluation of $S_{fi}$.
It is irrelevant if we are interested only in the absolute
value $|S_{fi}|$.
}

Following similar steps, one can show that in general
the Green function ${\cal G}$ appears as an intermediate
matrix element when one evaluates a transition amplitude
involving contributions from the quark-antiquark boundstates using 
time-ordered perturbation theory.

\section{\protect{Absence of Boundstate Poles in Eq.~(\ref{state})}}
\label{appb}

\begin{figure}[tbp]
  \begin{center}
    \includegraphics[width=10cm]{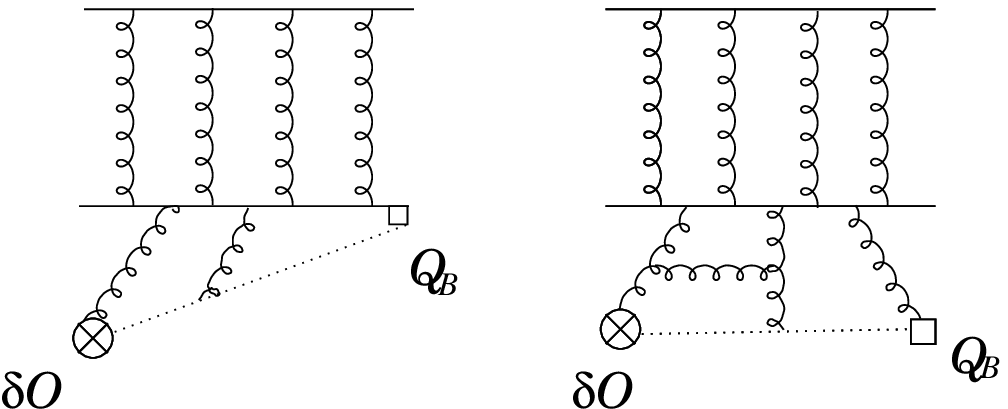}\\
    \vspace{8mm}
    \caption{\small \label{figc1}
Typical diagrams which have power countings 
$\beta^{n_0} \, (\alpha_S/\beta)^n$ for $n \gg n_0$.
The ghost is connected with the 
uncrossed ladders of $q\bar{q}$ with a finite number 
[$\simlt {\cal O}(n_0)$] of lines.
    }
  \end{center}
\end{figure}

We show that the state given by eq.~(\ref{state}) 
cannot accomodate a pole which is degenerate with any of the 
quark-antiquark boundstate poles,
$(P^0-M_\nu+i\epsilon)^{-1}$.
We first note that $Q_B$ and $\delta O$ have the ghost number 
$+1$ and $-1$, respectively.
Suppose this state contains some of these boundstate poles.
Then, 
the matrix element composed of this state should have a power counting 
in terms of $\alpha_S$ and $\beta$ as
\bea
&&
\bra{\vec{p},-\vec{p},\lambda,\bar{\lambda}} 
\delta O \,
\frac{1}{P^0-H+i\epsilon} \, i Q_B \,
\ket{\vec{q},-\vec{q},\lambda',\bar{\lambda}'} 
\nonumber \\ &&
~~~~~ ~~~~~
~~~~~ ~~~~~
\sim \beta^{n_0} \sum_n c_n \, (\alpha_S/\beta)^n \times
\biggl[ 1 + {\cal O}(1/c) \biggr] 
\eea
for some $n_0$, since 
$P^0 - M_\nu   = [ 1 - {C_F^2(\alpha_S/\beta)^2}/{4l^2} ] \, m \beta^2 $
at leading order.
It is known that the diagrams which can have the leading
power counting $(\alpha_S/\beta)^n$ are only the uncrossed ladder diagrams;
see Section~\ref{s3}.
Therefore the diagrams which can contribute to 
$\beta^{n_0} \, (\alpha_S/\beta)^n$
for $n \gg n_0$ are only those diagrams where a ghost is 
connected to one of the 
uncrossed ladders of $q\bar{q}$ with a finite number 
[$\simlt {\cal O}(n_0)$] of lines; see Fig.~\ref{figc1}.\footnote{
We discard the diagrams without cross talks between
$q\bar{q}$ and ghost sectors; see Section~\ref{s7}.
}
After integrating over the loop momenta, there remains no pole
in the $P^0$-dependence of the sum of the diagrams, in 
the same way that a usual one-loop diagram does not exhibit a pole
but rather contains branch point(s);
cf.\ eq.~(\ref{nlocharge}).

We may restate it differently.
If a ghost and a nonrelativistic 
$q\bar{q}$ pair should constitute a boundstate,
intuitively the sum of the ladder diagrams with multiple gluon exchanges
between the ghost and $q\bar{q}$ pair may exhibit a boundstate pole.
Since the coupling of 
ghost and gluon is suppressed by powers of $\beta$,
the binding energy of the boundstate should scale differently from
(have more powers of $\alpha_S$ than) the Coulomb binding energies
(if the boundstate should exist at all).

\end{appendix}

\newpage

\def\app#1#2#3{{\it Acta~Phys.~Polonica~}{\bf B #1} (#2) #3}
\def\apa#1#2#3{{\it Acta Physica Austriaca~}{\bf#1} (#2) #3}
\def\npb#1#2#3{{\it Nucl.~Phys.~}{\bf B #1} (#2) #3}
\def\plb#1#2#3{{\it Phys.~Lett.~}{\bf B #1} (#2) #3}
\def\prd#1#2#3{{\it Phys.~Rev.~}{\bf D #1} (#2) #3}
\def\pR#1#2#3{{\it Phys.~Rev.~}{\bf #1} (#2) #3}
\def\prl#1#2#3{{\it Phys.~Rev.~Lett.~}{\bf #1} (#2) #3}
\def\sovnp#1#2#3{{\it Sov.~J.~Nucl.~Phys.~}{\bf #1} (#2) #3}
\def\yadfiz#1#2#3{{\it Yad.~Fiz.~}{\bf #1} (#2) #3}
\def\jetp#1#2#3{{\it JETP~Lett.~}{\bf #1} (#2) #3}
\def\zpc#1#2#3{{\it Z.~Phys.~}{\bf C #1} (#2) #3}


\begin{thebibliography}{99}

\bibitem{cl} 
  W. Caswell and G. Lepage, 
  Phys.~Lett.~{\bf B167}, 437 (1986).

\bibitem{bbl}
  G. Bodwin, E. Braaten and G. Lepage, 
  Phys.~Rev.~{\bf D51}, 1125 (1995);  
  Erratum {\it ibid.} {\bf D55}, 5853 (1997).

\bibitem{labelle} 
  P. Labelle, 
  Phys.~Rev.~{\bf D58}, 093013 (1998). 

\bibitem{lm}
  M.~Luke and A.~Manohar, 
  Phys.~Rev.~{\bf D55}, 4129 (1997).

\bibitem{gr}
  B. Grinstein and I. Rothstein,
  Phys.~Rev.~{\bf D57}, 78 (1998). 

\bibitem{ls}
  M.~Luke and M.~Savage,
  Phys.~Rev.~{\bf D57}, 413 (1998).

\bibitem{pNRQCD} 
  A. Pineda and J. Soto, 
  Nucl.~Phys.~{\bf B}~(Proc. Suppl.)~{\bf 64}, 428 (1998);
  Phys.~Rev.~{\bf D59}, 016005 (1998).

\bibitem{bs} 
  M. Beneke and V. Smirnov, 
  Nucl.~Phys.~{\bf B522} 321 (1998). 

\bibitem{griesshammer}
  H. Griesshammer,
  Phys.~Rev.~{\bf D58}, 094027 (1998).

\bibitem{lmr}
  M.~Luke, A.~Manohar and I.~Rothstein,
  hep-ph/9910209.

\bibitem{renormalon}
  A.~Hoang, M.~Smith, T.~Stelzer and S.~Willenbrock, 
  Phys.~Rev.~{\bf D59}, 114014 (1999); 
  M.~Beneke, 
  {Phys.~Lett.}~{\bf B434}, 115 (1998).  

\bibitem{kn} 
  T. Kinoshita and M. Nio,
  Phys.~Rev.~{\bf D53}, 4909 (1996); Phys.~Rev.~{\bf D55}, 7267 (1997).

\bibitem{qed}
  G.~Adkins, R.~Fall and P.~Mitrikov, 
  Phys.~Rev.~Lett.~{\bf 79}, 3383 (1997);
  A.~Hoang, P.~Labelle and S.~Zebarjad,
  Phys.~Rev.~Lett.~{\bf 79}, 3387 (1997);
  hep-ph/9909495.

\bibitem{cmy} 
  A. Czarnecki, K. Melnikov and A. Yelkhovsky,
  Phys.~Rev.~Lett.~{\bf 82}, 311 (1999);  
  Phys.~Rev.~{\bf A59}, 4316 (1999);
  Phys.~Rev.~Lett.~{\bf 83}, 1135 (1999).

\bibitem{qcdpot}
  M.~Peter, Phys.~Rev.~Lett.~{\bf 78}, 602 (1997); 
  Nucl.~Phys.~{\bf B501} 471 (1997);
  Y.~Schr\"oder, {Phys.~Lett.}~{\bf B447}, 321 (1999).  

\bibitem{py}
  A. Pineda and  F. Yndur{\'a}in, 
  Phys.~Rev.~{\bf D58}, 094022 (1998);
  hep-ph/9812371.

\bibitem{NNLO} 
  A. Hoang and T. Teubner, Phys.~Rev.~{\bf D58}, 114023 (1998);
  K.~Melnikov and A.~Yelkhovsky, Nucl.~Phys.~{\bf B528}, 59 (1998).

\bibitem{upsilon}
  A. Hoang,
  Phys.~Rev.~{\bf D59}, 014039 (1999).

\bibitem{analytic}
  K.~Melnikov and A.~Yelkhovsky, 
  Phys.~Rev.~{\bf D59}, 114009 (1999);
  A.~Penin and A.~Pivovarov, Nucl.~Phys.~{\bf B549}, 217 (1999);
  hep-ph/9904278.

\bibitem{yakovlev}
  O.~Yakovlev, hep-ph/9808463.

\bibitem{bpsv} 
  N.~Brambilla, A.~Pineda, J.~Soto and A.~Vairo, 
  Phys.~Rev.~{\bf D60}, 091502 (1999);
  hep-ph/9907240; hep-ph/9910238.

\bibitem{bss} 
  M. Beneke, A. Signer and V. Smirnov, 
  hep-ph/9903260.

\bibitem{momdist}
  T.~Nagano, A.~Ota and Y.~Sumino, hep-ph/9903498;
  A.~Hoang and T.~Teubner, hep-ph/9904468.

\bibitem{kp} B. Kniehl and A. Penin,  hep-ph/9907489.

\bibitem{by}
  G. Bodwin and D. Yennie, 
  Phys.~Rep.~{\bf 43}, 267 (1978).

\bibitem{love}
  S.~Love,
  Ann.~of~Phys.~{\bf 113}, 153 (1978).

\bibitem{ffh}
  G.~Feldman, T.~Fulton and D.~Heckathorn, 
  {Nucl.~Phys.}~{\bf B167}, 364 (1980);  {Nucl.~Phys.}~{\bf B174}, 89 (1980).

\bibitem{lw}
  M. Luscher and P. Weisz,
  Comm.~Math.~Phys.~{\bf 97}, 59 (1985); 
  Erratum {\it ibid.}~{\bf 98}, 433 (1985). 

\bibitem{sf}
  S. Scherer and H. Fearing,
  Phys.~Rev.~{\bf D52}, 6445 (1995).

\bibitem{canonical}
  T.~Kugo and I.~Ojima,
  Phys.~Lett.~{\bf 73B}, 459 (1978);
  Prog.~Theor.~Phys.~Suppl.~{\bf 66}, 1 (1979);
  N.~Nakanishi and I.~Ojima,
  ``Covariant Operator Formalism of Gauge Theories and
    Quantum Gravity'', 
  (World Scientific Lecture Notes in Physics Vol.~27, 1990).

\bibitem{lmt}
  D.~Luri\'e, A.~MacFarlane and Y.~Takahashi,
  Phys.~Rev.~{\bf 140B}, 1091 (1965).

\bibitem{braun}
  M.~Braun,
  Sov.~Phys.~JETP~{\bf 27}, 652 (1968).

\bibitem{ap}
  T.~Appelquist and H.~Politzer,
  Phys.~Rev.~Lett.~{\bf 34}, 43 (1975);
  Phys.~Rev.~{\bf D12}, 1404 (1975).

\bibitem{text}
  F.~Halzen and A.~Martin,
  ``Quarks \& Leptons'', (John Wiley \& Sons, 1984).

\bibitem{heckathorn}
  D.~Heckathorn, 
  {Nucl.~Phys.}~{\bf B156}, 328 (1979).

\bibitem{fsi} 
  Y. Sumino, PhD thesis, University of Tokyo 1993 (unpublished);
  M.~Peter and Y.~Sumino, Phys.~Rev.~{\bf D57}, 6912 (1998).

\bibitem{lattice1}
  B. Thacker and G. Lepage, Phys.~Rev.~{\bf D43}, 196 (1991);
  G.~Lepage, L.~Magnea and C. Nakhleh,
  Phys.~Rev.~{\bf D46}, 4052 (1992).

\bibitem{lattice2}
  A.~Khan, J.~Shigemitsu, S.~Collins, C.~Davies, C.~Morningstar
  and J.~Sloan,   Phys.~Rev.~{\bf D56}, 7012 (1997);
  K.~Ishikawa, H.~Matsufuru, T.~Onogi, N.~Yamada and S.~Hashimoto,
  Phys.~Rev.~{\bf D56}, 7028 (1997).

\bibitem{sumino}
Y. Sumino, K. Fujii, K. Hagiwara, H. Murayama, and
C.-K. Ng,  Phys.~Rev. {\bf D47}, 56 (1993);
M. Je\.zabek, J.H. K\"uhn and T. Teubner, Z. Phys. {\bf C56}, 653 (1992);
M. Je\.zabek and T. Teubner, Z. Phys. {\bf C59}, 669 (1993).

\end{thebibliography}
\end{document}